\documentclass[twocolumn]{aastex631}
\usepackage{amssymb,mathtools}
\usepackage{CJK}
\shorttitle{SCUBA-2 COSMOS 450 $\mu \mathrm{m}$}
\shortauthors{Gao et al.}
\graphicspath{{./}}

\begin{document}
\begin{CJK*}{UTF8}{bsmi}
\title{SCUBA-2 Ultra Deep Imaging EAO Survey (STUDIES). V. Confusion-limited Submillimeter Galaxy Number Counts at 450 $\micron$ and Data Release for the COSMOS Field}

\author[0000-0003-1262-7719]{Zhen-Kai Gao (高振凱)}
\affiliation{Academia Sinica Institute of Astronomy and Astrophysics (ASIAA), No. 1, Section 4, Roosevelt Rd., Taipei 10617, Taiwan}
\affiliation{Graduate Institute of Astronomy, National Central University, 300 Zhongda Road, Zhongli, Taoyuan 32001, Taiwan}

\author[0000-0003-1213-9360]{Chen-Fatt Lim (林征發)}
\affiliation{Academia Sinica Institute of Astronomy and Astrophysics (ASIAA), No. 1, Section 4, Roosevelt Rd., Taipei 10617, Taiwan}

\author[0000-0003-2588-1265]{Wei-Hao Wang (王為豪)}
\affiliation{Academia Sinica Institute of Astronomy and Astrophysics (ASIAA), No. 1, Section 4, Roosevelt Rd., Taipei 10617, Taiwan}

\author[0000-0002-3805-0789]{Chian-Chou Chen (陳建州)}
\affiliation{Academia Sinica Institute of Astronomy and Astrophysics (ASIAA), No. 1, Section 4, Roosevelt Rd., Taipei 10617, Taiwan}

\author[0000-0003-3037-257X]{Ian Smail}
\affiliation{Centre for Extragalactic Astronomy, Department of Physics, Durham University, South Road, Durham DH1 3LE, UK}

\author[0000-0002-8487-3153]{Scott C.\ Chapman}
\affiliation{Department of Physics and Atmospheric Science, Dalhousie University, Halifax, Halifax, NS B3H 3J5, Canada}

\author[0000-0003-3728-9912]{Xian Zhong Zheng}
\affiliation{Purple Mountain Observatory and Key Laboratory for Radio Astronomy, Chinese Academy of Sciences, 10 Yuanhua Road, Nanjing 210023, China}
\affiliation{School of Astronomy and Space Science, University of Science and Technology of China, Hefei 230026, Anhui, China}

\author[0000-0002-4179-2628]{Hyunjin Shim}
\affiliation{Department of Earth Science Education, Kyungpook National University, 80 Daehak-ro, Buk-gu, Daegu 41566, Republic of Korea}

\author[0000-0002-2993-1576]{Tadayuki Kodama}
\affiliation{Astronomical Institute, Tohoku University, 6-3, Aramaki, Aoba, Sendai, Miyagi, 980-8578, Japan}

\author[0000-0003-3139-2724]{Yiping Ao}
\affiliation{Purple Mountain Observatory and Key Laboratory for Radio Astronomy, Chinese Academy of Sciences, 10 Yuanhua Road, Nanjing 210023, China}
\affiliation{School of Astronomy and Space Science, University of Science and Technology of China, Hefei 230026, Anhui, China}

\author[0009-0001-1744-1246]{Siou-Yu Chang (張修瑜)}
\affiliation{Institute of Astronomy and Department of Physics, National Tsing Hua University, Hsinchu 30013, Taiwan}

\author[0000-0002-9548-5033]{David L. Clements}
\affiliation{Imperial College London, Blackett Laboratory, Prince Consort Road, London SW7 2AZ, UK}

\author{James S. Dunlop}
\affiliation{SUPA, Institute for Astronomy, University of Edinburgh, Royal Observatory, Edinburgh EH9 3HJ, UK}

\author[0000-0001-6947-5846]{Luis C. Ho}
\affiliation{Kavli Institute for Astronomy and Astrophysics, Peking University, Beijing 100871, China}
\affiliation{Department of Astronomy, School of Physics, Peking University, Beijing 100871, China}

\author[0000-0003-0381-562X]{Yun-Hsin Hsu (徐允心)}
\affiliation{Academia Sinica Institute of Astronomy and Astrophysics (ASIAA), No. 1, Section 4, Roosevelt Rd., Taipei 10617, Taiwan}
\affiliation{Department of Philosophy, National Taiwan University, No. 1, Section 4, Roosevelt Rd., Taipei 10617, Taiwan}
\affiliation{Current address: University Observatory, Faculty of Physics, Ludwig-Maximilians-Universität München, Scheinerstr. 1, 81679 Munich, Germany}

\author[0000-0002-3658-0903]{Chorng-Yuan Hwang}
\affiliation{Graduate Institute of Astronomy, National Central University, 300 Zhongda Road, Zhongli, Taoyuan 32001, Taiwan}

\author[0000-0003-3428-7612]{Ho Seong Hwang}
\affiliation{Astronomy Program, Department of Physics and Astronomy, Seoul National University, 1 Gwanak-ro, Gwanak-gu, Seoul 08826, Republic of Korea}
\affiliation{SNU Astronomy Research Center, Seoul National University, 1 Gwanak-ro, Gwanak-gu, Seoul 08826, Republic of Korea}

\author[0000-0001-5785-1154]{M. P. Koprowski}
\affiliation{Institute of Astronomy, Faculty of Physics, Astronomy and Informatics, Nicolaus Copernicus University, Grudziadzka 5, 87-100 Torun, Poland}

\author[0000-0002-6878-9840]{Douglas Scott}
\affiliation{Department of Physics and Astronomy, University of British Columbia, 6224 Agricultural Road, Vancouver, BC V6T 1Z1, Canada}

\author[0000-0002-0517-7943]{Stephen Serjeant}
\affiliation{School of Physical Sciences, The Open University, Milton Keynes, MK7 6AA, UK}

\author[0000-0002-3531-7863]{Yoshiki Toba}
\affiliation{National Astronomical Observatory of Japan, 2-21-1 Osawa, Mitaka, Tokyo 181-8588, Japan}
\affiliation{Academia Sinica Institute of Astronomy and Astrophysics (ASIAA), No. 1, Section 4, Roosevelt Rd., Taipei 10617, Taiwan}
\affiliation{Research Center for Space and Cosmic Evolution, Ehime University, 2-5 Bunkyo-cho, Matsuyama, Ehime 790-8577, Japan}

\author[0000-0001-9708-3831]{Sheona A. Urquhart}
\affiliation{School of Physical Sciences, The Open University, Milton Keynes, MK7 6AA, UK}



\begin{abstract}

We present confusion-limited SCUBA-2 450-$\micron$ observations in the COSMOS-CANDELS region as part of the JCMT Large Program, SCUBA-2 Ultra Deep Imaging EAO Survey (STUDIES). Our maps at 450 and 850\,$\micron$ cover an area of 450\,arcmin$^2$. We achieved instrumental noise levels of $\sigma_{\mathrm{450}}=$ 0.59\,mJy\,beam$^{-1}$ and $\sigma_{\mathrm{850}}=$ 0.09\,mJy\,beam$^{-1}$ in the deepest area of each map. The corresponding confusion noise levels are estimated to be 0.65 and 0.36\,mJy\,beam$^{-1}$. Above the 4 (3.5) $\sigma$ threshold, we detected 360 (479) sources at 450\,$\micron$ and 237 (314) sources at 850\,$\micron$. We derive the deepest blank-field number counts at 450\,$\micron$, covering the flux-density range of 2 to 43\,mJy. These are in agreement with other SCUBA-2 blank-field and lensing-cluster observations, but are lower than various model counts.
We compare the counts with those in other fields and find that the field-to-field variance observed at 450\,$\micron$ at the $R=6\arcmin$ scale is consistent with Poisson noise, so there is no evidence of strong 2-D clustering at this scale.
Additionally, we derive the integrated surface brightness at 450\,$\micron$ down to 2.1\,mJy to be $57.3^{+1.0}_{-6.2}$~Jy\,deg$^{-2}$, contributing to (41$\pm$4)\% of the 450-$\micron$ extragalactic background light (EBL) measured by \emph{COBE} and \emph{Planck}. Our results suggest that the 450-$\micron$ EBL may be fully resolved at $0.08^{+0.09}_{-0.08}$~mJy, which extremely deep lensing-cluster observations and next-generation submillimeter instruments with large aperture sizes may be able to achieve.

\end{abstract}



\section{Introduction} \label{sec:intro}

The result that the portions of the extragalactic background light (EBL) in the infrared and in the optical are comparable \citep{pugetTentativeDetectionCosmic1996,fixsenSpectrumExtragalacticFar1998,coorayExtragalacticBackgroundLight2016,hillSpectrumUniverse2018} implies that approximately half of the cosmic star-formation activity is obscured by dust \citep[see a review in][]{madauCosmicStarFormationHistory2014}. Dust absorbs ultraviolet photons and re-emits them at mid-infrared to millimeter wavelengths. The spectral energy distributions (SEDs) of galaxies can therefore be significantly reshaped by the presence of dust. Since the advent of the Submillimeter Common User Bolometer Array \citep[SCUBA;][]{hollandSCUBACommonuserSubmillimetre1999}, mounted on the 15-meter James Clerk Maxwell Telescope (JCMT), the dust-reprocessed portion of the EBL has been largely resolved into so-called submillimeter galaxies \citep[SMGs,][]{smailDeepSubmillimeterSurvey1997,bargerSubmillimetrewavelengthDetectionDusty1998,hughesHighredshiftStarFormation1998,ealesCanadaUKDeep1999}, opening a new era of studying galaxies in the submillimeter regime.

Over the past two decades, numerous studies of SMGs selected at 850 $\micron$ and millimeter wavelengths made with bolometer array cameras \citep[see reviews in][]{blainClusteringSubmillimeterSelected2004,caseyAreDustyGalaxies2014} and interferometric observations \citep[e.g.,][]{wangGOODS8505Galaxy2007,youngerEvidencePopulationHighRedshift2007,youngerClarifyingNatureBrightest2008,cowieAccuratePositionHDF2009,hodgeALMASurveySubmillimeter2013,karimALMASurveySubmillimetre2013,vieiraDustyStarburstGalaxies2013,cowieSubmillimeterPerspectiveGOODS2017,cowieSubmillimeterPerspectiveGOODS2018,stachALMASurveySCUBA22019,simpsonALMASurveyBrightest2020,chenALMASpectroscopicSurvey2022,cowieSubmillimeterSurveyFaint2022,chenALMACALIXMultiband2023,fujimotoALMALensingCluster2023} have expanded our understanding of this dusty galaxy population. We now know that these SMGs are gas-rich \citep{greveInterferometricCOSurvey2005,tacconiHighResolutionMillimeter2006,bothwellSurveyMolecularGas2013,birkinALMANOEMASurvey2021} and have high star-formation rates (SFRs) of $100\textrm{--}1000~\mathrm{M_{\odot}}~\mathrm{yr}^{-1}$ \citep{bargerTHEREMAXIMUMSTAR2014,swinbankALMASurveySubmillimetre2014a,shimMultiwavelengthProperties850mm2022}. Furthermore, they are massive ($\sim 10^{11}\;\mathrm{M_{\odot}}$) \citep{dyeSCUBAHAlfDegree2008,hainlineStellarMassContent2011,michalowskiStellarMassesSpecific2012a,smolcicPhysicalPropertiesOfz2015,koprowskiSCUBA2CosmologyLegacy2016,michalowskiSCUBA2CosmologyLegacy2017,dudzeviciuteALMASurveySCUBA22020a} and reside in halos that will eventually evolve into hosting massive elliptical galaxies at $z=0$ \citep{hickoxLABOCASurveyExtended2012,hildebrandtInferringMassSubmillimetre2013,chenFAINTSUBMILLIMETERGALAXIES2016,wilkinsonSCUBA2CosmologyLegacy2017,anMultiwavelengthPropertiesRadio2019,stachALMASurveySCUBA22021}. The redshift distribution of 850-$\micron$-selected SMGs peaks at $z=2\textrm{--}3$ \citep[e.g.,][]{bargerMappingEvolutionHighRedshift2000,chapmanMedianRedshiftGalaxies2003,simpsonALMASurveySubmillimeter2014,chenSCUBA2CosmologyLegacy2016,michalowskiSCUBA2CosmologyLegacy2017,zavalaSCUBA2CosmologyLegacy2018,dudzeviciuteALMASurveySCUBA22020a,reuterCompleteRedshiftDistribution2020,chenALMASpectroscopicSurvey2022}, which is close to the peak of the cosmic star-formation rate density \citep{madauCosmicStarFormationHistory2014,fermi-latcollaborationGammarayDeterminationUniverse2018,driverGAMAG10COSMOS3DHST2018,lopezfernandezCosmicEvolutionSpatially2018,sanchezSDSSIVMaNGAArchaeological2019,bellstedtGalaxyMassAssembly2020} and the peak of active galactic nuclei (AGN) activity \citep{schmidtSpectrscopicCCDSurveys1995,hasingerLuminositydependentEvolutionSoft2005,wallEvolutionSubmillimetreGalaxies2008a, assefMidIRXraySelectedQSO2011,uedaStandardPopulationSynthesis2014,airdEvolutionXrayLuminosity2015} at $z\sim2$. However, SMGs studied with single-dish telescopes are naturally limited by the effects of confusion, which only allows the detections of sources brighter than $S_{850} \sim 2$~mJy. These sources only comprise up to $1/3$ of the submillimeter EBL \citep{bargerResolvingSubmillimeterBackground1999,cowieFaintSubmillimeterCounts2002}. To reach a more complete picture of dusty galaxies that give rise to the submillimeter EBL, we need higher angular resolution to go beyond the 850-$\micron$ confusion limit and to detect the more typical members of the SMG population. While the Atacama Large Millimeter/submillimeter Array (ALMA) can provide the required resolution and sensitivity to detect fainter SMGs \citep[e.g.,][]{oteoALMACALFirstDualband2016,walterALMASpectroscopicSurvey2016,dunlopDeepALMAImage2017a,francoGOODSALMAMmGalaxy2018,hatsukadeALMATwentysixArcmin22018}, the small field of view of ALMA (FWHM$\,=17.3$~arcsec) limits sample sizes. 
Given the above challenges with 850-$\micron$ observations, we take another approach, which is to observe with the 450-$\micron$ channel of SCUBA-2 \citep{hollandSCUBA2100002013}. The 450-$\micron$ band probes closer to the peaks of the dust SEDs on galaxies at moderate redshifts of $z<4$ including the ``cosmic noon'' at $z\sim2$. In this paper, we will present our extremely deep SCUBA-2 450-$\micron$ imaging.

A major advantage of observing at 450~$\micron$ is the roughly two times higher angular resolution (7\farcs5 FWHM, as opposed to $14\arcsec$ at 850~$\micron$). This makes it possible to detect faint SMGs below the confusion limit at 850~$\micron$ that are selected at a wavelength closer to the peak of the EBL, which comprise the bulk of the EBL. The better positional accuracy provided by the higher resolution also means that counterpart identification would be less challenging, which can lead to multi-wavelength studies of larger samples of 450-$\micron$ SMGs \citep[e.g.,][]{roseboomSCUBA2CosmologyLegacy2013,changSCUBA2UltraDeep2018, limSCUBA2UltraDeep2020,dudzeviciuteTracingEvolutionDustobscured2021}.
Furthermore, because 450~$\micron$ is closer to the rest-frame peak of the dust SED, it can probe less luminous population that dominated the star formation rate density at $z\sim1$--2 \citep[][]{caseyCharacterizationSCUBA24502013,roseboomSCUBA2CosmologyLegacy2013, zavalaSCUBA2CosmologyLegacy2018, limSCUBA2UltraDeep2020,bargerSubmillimeterPerspectiveGOODS2022a}, to which 850-$\micron$ observations are less sensitive. Although the \emph{Herschel} Spectral and Photometric Imaging Receiver (SPIRE) provides deep 250-, 350-, and 500-$\micron$ imaging, the high confusion limits of 19, 18, and 15 mJy, respectively, \citep{nguyenHerMESSPIREConfusion2010} prevent us from going deeper and detecting the typical members of the dusty galaxy population. Therefore, 450-$\micron$ observations conducted by SCUBA-2 can help to close the gap between far-infrared and 850-$\micron$ observations.

Although the high angular resolution of SCUBA-2 at 450~$\micron$ is advantageous, observing at 450~$\micron$ is highly challenging and requires the best weather condition (the ``band 1'' condition, with the atmospheric opacity $\tau < 0.05$ at 225 GHz) on Maunakea. Because of the limited number of nights that fulfill the required band-1 weather condition, deep SCUBA-2 450-$\micron$ blank-field surveys were only conducted in three fields---the Cosmological Evolution Survey \citep[COSMOS,][]{caseyCharacterizationSCUBA24502013,geachSCUBA2CosmologyLegacy2013,wangSCUBA2UltraDeep2017}, the Extended Groth Strip \citep[EGS,][]{zavalaSCUBA2CosmologyLegacy2017}, and the Great Observatories Origins Deep Survey North and South \citep[GOODS-N and GOODS-S, also known as CDF-N and CDF-S,][]{cowieSubmillimeterPerspectiveGOODS2017}. The survey areas and central rms noise levels of the fields, except for COSMOS, are all around $100$~arcmin$^2$ and $>1$~mJy~beam$^{-1}$. To achieve a sensitivity comparable to the confusion noise of SCUBA-2 at 450~$\micron$ ($0.6$~mJy~beam$^{-1}$), we conducted a JCMT Large Program: the SCUBA-2 Ultra Deep Imaging EAO (East Asian Observatory) Survey \citep[STUDIES,][]{wangSCUBA2UltraDeep2017}, to image the centers of the COSMOS \citep{scovilleCosmicEvolutionSurvey2007} and Subaru/$\emph{XMM-Newton}$ Deep Survey \citep[SXDS, also known as UKIDSS/UDS,][]{furusawaSubaruXMMNewtonDeep2008,lawrenceUKIRTInfraredDeep2007} fields with extreme depths at 450~$\micron$ and an area of $\sim180$~arcmin$^2$ per pointing (two pointings in COSMOS and one in SXDS). The 330 hours of observations for COSMOS (hereafter STUDIES-COSMOS, JCMT project ID: M16AL006) were completed in 2020, and the 320 hours of observations for SXDS (hereafter STUDIES-SXDS, JCMT project ID: M17BL009) are still ongoing. Our early results include the detection of a $z=3.7$ ``passive'' galaxy at 450~$\micron$ \citep{simpsonImperfectlyPassiveNature2017}, intermediate-depth 450-$\micron$ number counts \citep{wangSCUBA2UltraDeep2017}, rest-frame optical morphologies of 450-$\micron$ and 850-$\micron$ SMGs \citep{changSCUBA2UltraDeep2018}, far-infrared luminosity functions \citep{limSCUBA2UltraDeep2020}, clustering of machine-learning-selected 450-$\micron$ SMGs \citep{limSCUBA2UltraDeep2020a}, comparison of the physical properties of 450-$\micron$ and 850-$\micron$ SMGs \citep{dudzeviciuteTracingEvolutionDustobscured2021}, and a strongly lensed SMGs discovered using James Webb Space Telescope (JWST) Mid-Infrared Instrument (MIRI) at 7.7 $\micron$ \citep{pearsonLargePopulationStrongly2024}. These works show that we can reach sizeable samples of more than $300$ 450-$\micron$ SMGs and detect SMGs in the star-forming main sequence, even with preliminary data.  

In this paper, we present the first confusion-limited 450-$\micron$ image made with STUDIES-COSMOS data and archival data, the deepest 450-$\micron$ blank-field number counts, and their contributions to the 450-$\micron$ EBL. We publicly release our final STUDIES-COSMOS maps and catalogs at both 450~$\micron$ and 850~$\micron$.\footnote{\url{http://group.asiaa.sinica.edu.tw/whwang/studies/cosmos\_final/}} Additionally, we present preliminary 450-$\micron$ number counts from 100 hours of integration of STUDIES-SXDS. We compare the counts derived from various fields and study the field-to-field variance inferred from the data and from models. We also compare the EBL contribution of the resolved 450-$\micron$ sources with the EBL measurements from \emph{Cosmic Background Explorer} (\emph{COBE}) Far Infrared Absolute Spectrophotometer (FIRAS) and \emph{Planck} High Frequency Instrument (HFI).
In Section~\ref{sec:data}, we describe the SCUBA-2 observations and data reduction, and present the 450-$\micron$ and 850-$\micron$ images and catalogs. In Section~\ref{sec:number_counts}, we compute the raw number counts and use simulations to correct for observational biases to derive the intrinsic counts. In Section~\ref{sec:discussion}, we compare our number counts with those from previous observations and models, estimate the variance between fields, and calculate the contributions to the 450-$\micron$ EBL. We summarize our results in Section~\ref{sec:summary}. In Appendix~\ref{apd:confusion_limit}, we estimate the confusion limits at 450 and 850~$\micron$ using the number counts. In Appendix~\ref{apd:fcf_corr_factor}, we describe the flux density correction factors applied to previous observations for our comparison of number counts. We adopt a $\Lambda$CDM cosmology with $\Omega_{\Lambda}=0.7$, $\Omega_{\mathrm{m}}=0.3$, and $H_{0}=70~\mathrm{km~s^{-1}~Mpc^{-1}}$, unless otherwise stated.

\section{Observations and Data Reduction} \label{sec:data}

\subsection{STUDIES 450 \texorpdfstring{$\micron$}{micron} Observations}

\begin{figure*}
\epsscale{1.7}
\plottwo{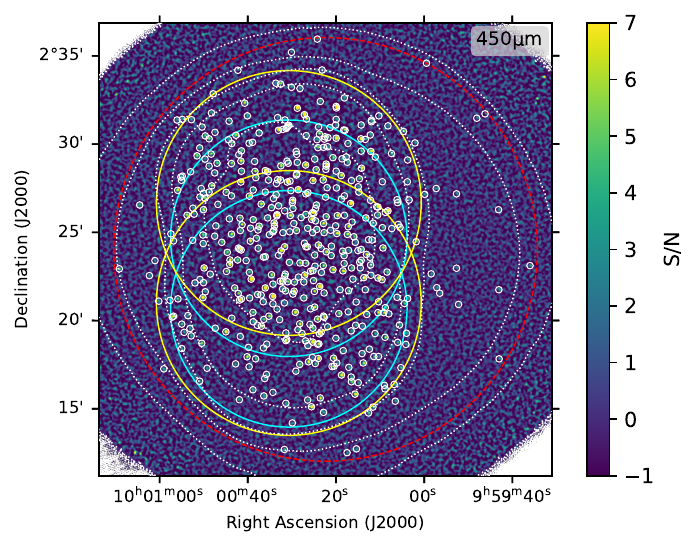}{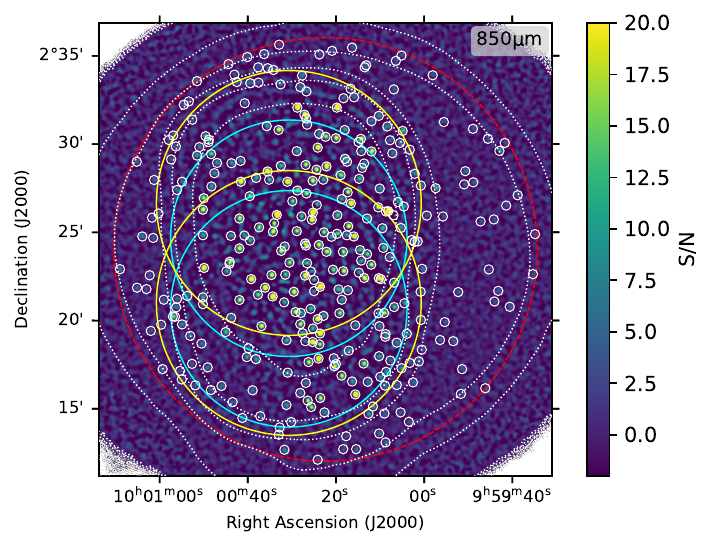}
\caption{The final STUDIES-COSMOS 450-$\micron$ and 850-$\micron$ S/N maps. Each combined map is composed of the maps from \citet[][the whole coverage]{caseyCharacterizationSCUBA24502013}, STUDIES-COSMOS (this work, yellow circles), and S2CLS (cyan circles). The white circles mark 479 (450~$\micron$) and 321 (850~$\micron$) $>$3.5$\sigma$ sources identified within the $R=12\arcmin$ region (red dashed circles). The white dotted contours show the noise levels of, innermost, 1~(0.2)~mJy with a multiplicative step of 2 for 450~(850)~$\micron$ (e.g., 1, 2, 4, etc.).}
\label{fig:coverage_map}
\end{figure*}

The STUDIES-COSMOS program \citep[][project ID: M16AL006]{wangSCUBA2UltraDeep2017} has accomplished the first confusion-limited observations at 450~$\micron$, in the COSMOS field of the Cosmic Assembly Near-infrared Deep Extragalactic Legacy Survey \citep[CANDELS,][]{groginCANDELSCosmicAssembly2011,koekemoerCANDELSCosmicAssembly2011}. The primary pointing center (541 scans) is at RA $=10^{\mathrm{h}}$00$^{\mathrm{m}}$30\fs7 and Dec $=+02\degr26\arcmin40\arcsec$, while the secondary pointing center (71 scans) is at RA $=10^{\mathrm{h}}$00$^{\mathrm{m}}$30\fs7 and Dec $=+02\degr21\arcmin00\arcsec$ to cover the whole CANDELS footprint. The observations were carried out between 2015 December 30 and 2020 June 15, with a total on-sky integration time reaching 314 hours with the SCUBA-2 instrument \citep{hollandSCUBA2100002013} mounted on the JCMT.
The observations were performed only under the best submillimeter weather conditions (``Band-1'' weather, $\tau_{\mathrm{225GHz}}<0.05$) to maximize the atmospheric transmission ($>28\%$ at 450~$\micron$ and $>82\%$ at 850~\micron\footnote{\url{https://www.eaobservatory.org/jcmt/observing/weather-bands/}}) especially for the 450 $\micron$ band.  This makes the observations very challenging because typically only 15\% to 20\% of observing time in winters falls in Band 1.
To obtain the deepest map, we chose the ``CV Daisy'' scan pattern, which keeps the pointing center always covered by one of the four SCUBA-2 sub-arrays. Additionally, we included minor dithering in the observations. By using this scan pattern along with the dithering, this produces a circular map with a radius of 7.5 arcmin. The coverage is shown by the yellow circles in Figure \ref{fig:coverage_map}.

\subsection{Archival Data}

Besides STUDIES, we combine data from two other sets of archival SCUBA-2 programs that overlap with our data to increase the depth and area of the map. The first is the extremely deep map observed using the ``CV Daisy'' scan pattern by the SCUBA-2 Cosmology Legacy Survey \citep[S2CLS,][project ID: MJLSC01]{geachSCUBA2CosmologyLegacy2013}. S2CLS has two pointing centers in the COSMOS-CANDELS field (cyan circles in Figure \ref{fig:coverage_map}). They have the same RA as STUDIES, but have Dec that are 2$\arcmin$ and 6$\arcmin$ further south than the primary pointing center of STUDIES. With this partially overlapping map we can increase the depth and area to help detect fainter sources and increase the sample size in all flux density ranges. The second set of observations that we include are the shallower but wider mapping observed using the ``PONG'' scan pattern by \citet[][project ID: M11BH11A, M12AH11A, and M12BH21A]{caseyCharacterizationSCUBA24502013}, which has a mapping center of RA $=10^{\mathrm{h}}$00$^{\mathrm{m}}$28\fs0 and Dec $=+02\degr24\arcmin00\arcsec$. This provides the wide area coverage shown in Figure~\ref{fig:coverage_map}, which includes both the STUDIES and S2CLS regions. In areas where this map overlaps with STUDIES and S2CLS, it does not significantly increase the overall depth, but it does improve the noisy outskirts. This additional wide and shallow coverage yields the detection of approximately a dozen extra bright sources. The additional data from S2CLS and from \citet{caseyCharacterizationSCUBA24502013} contain integration times of 155 and 33 hours, respectively, bringing the total integration time to 502 hours.

\begin{figure}
\epsscale{1.17}
\plotone{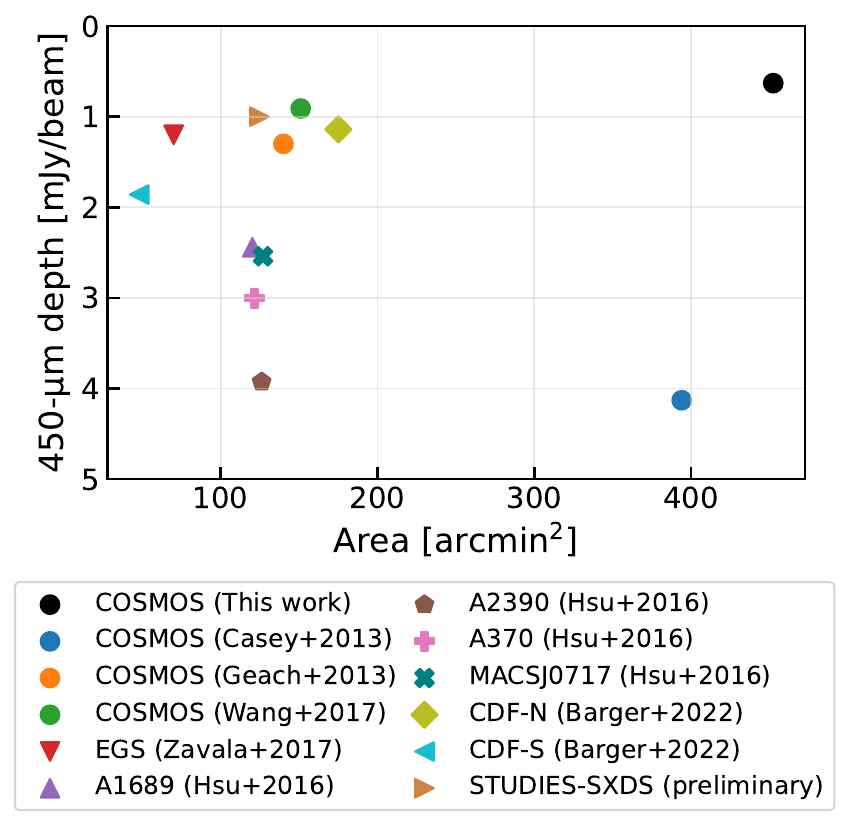}
\caption{Depth and area of various SCUBA-2 450-$\micron$ surveys. Those with areas less than 200~arcmin$^2$ were made with the ``CV DAISY'' scan mode, while wider ones were made with (or including) the ``PONG'' scan mode. The depth refers to the instrumental noise at the map center. \citet{hsuHawaiiSCUBA2Lensing2016} presents the mean depth of the selected map area. To convert the mean depths to the central depths, we apply a factor of 0.55, estimated from a noise map of the ``CV DAISY'' scan.}
\label{fig:depth_area}
\end{figure}

\begin{figure}
\epsscale{1.17}
\plotone{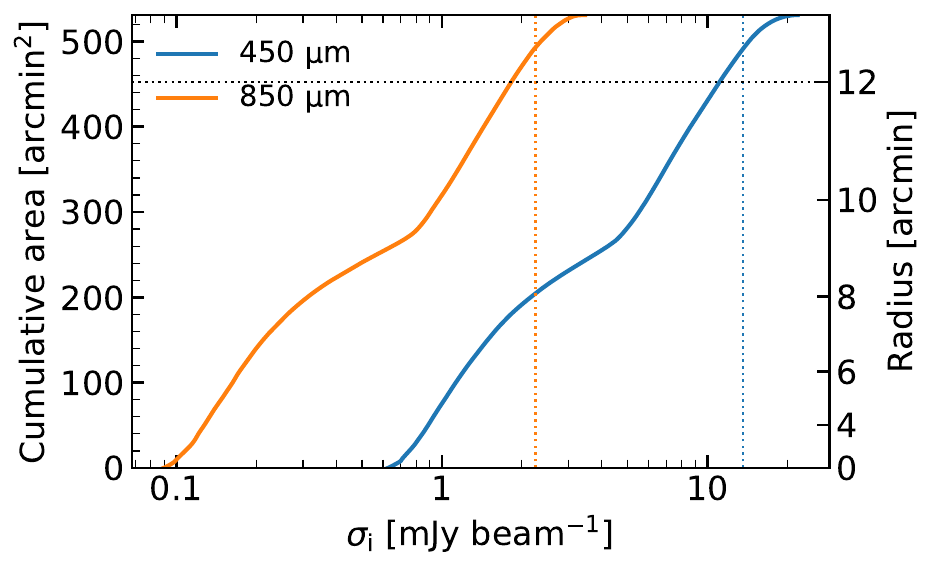}
\caption{Cumulative area of our 450-$\micron$ (blue curve) and 850-$\micron$ (orange curve) maps as a function of instrumental noise level. The dotted black horizontal line marks the selected $R=12\arcmin$ regions (red dashed circles in Figure~\ref{fig:coverage_map}), and the two dotted vertical lines mark the highest instrumental noise levels in the two regions.
Because the sensitivity distributions of the maps are not circularly symmetric, some pixels with higher instrumental noise levels are located within the $R=12\arcmin$ region. This causes the shifts between the intersections of the dotted lines (vertical and horizontal) and the solid curve for each waveband.}
\label{fig:cumulative_noise_area}
\end{figure}

The final combined S/N maps at 450~$\micron$ and 850~$\micron$ are shown in Figure \ref{fig:coverage_map} (i.e., whole coverage). The region we use in this work is centered at RA $=10^{\mathrm{h}}$00$^{\mathrm{m}}$22\fs36 and Dec $=+02\degr24\arcmin02\farcs00$ with a radius of $12\arcmin$ (red dashed circles in Figure \ref{fig:coverage_map}). At the center of the deep area, the rms instrumental noise\footnote{The ``instrumental noise'' term encompasses more than strictly instrument noise. Specifically, it folds in additional variance from atmospheric fluctuations and other observational uncertainties. However, this aggregated noise term persists in its ``instrumental'' naming by convention, to maintain consistent terminology with the literature where this noise combination is not always explicitly spelled out.} is 0.59~mJy at 450~$\micron$ and 0.09~mJy at 850~$\micron$ (see Section~\ref{section_noise} for more details), making this the deepest SCUBA-2 map with moderate area coverage (450~arcmin$^2$).
In Figure \ref{fig:depth_area} we show the depth and area of our combined 450-$\micron$ map and compare it with other SCUBA-2 450-$\micron$ surveys. Also for comparison, the Herschel Astrophysical Terahertz Large Area Survey \citep[H-ATLAS,][]{valianteHerschelATLASData2016} has rms noise levels of approximately $10$~mJy at 350 and 500~$\micron$ and an area coverage of 162~deg$^2$. The dramatically different sensitivities and areas of the SCUBA-2 and \emph{Herschel} surveys make them highly complementary to each other. In Figure~\ref{fig:cumulative_noise_area} we show the cumulative area as a function of the instrumental noise cut at 450~$\micron$ and 850~$\micron$. The poorest sensitivities at the outskirts of the $R=12\arcmin$ circles are 13.3~mJy~beam$^{-1}$ at 450~$\micron$ and 2.17~mJy~beam$^{-1}$ at 850~$\micron$.

\subsection{Data Reduction} \label{data_reduction}

Our data reduction consists of the following procedures: (1) making maps from raw data; (2) selecting maps taken under good weather conditions; (3) converting to flux density units; (4) mosaicking all scans; (5) removing bad pixels; and (6) applying matched filters to the maps. We adopt the Starlink package (\citealp{currieStarlinkSoftware20132014} version 2018A) for these tasks.\\

(1) We applied the \texttt{makemap} command provided by the Sub-Millimeter Common User Reduction Facility \citep[SMURF,][]{chapinSCUBA2IterativeMapmaking2013} to all 30-minute time streams with the standard ``blank-field'' configuration file and with the pixel scale set to one arcsec per pixel. We call this reduced 30-minute time stream a ``scan.''\\

(2) After the map-making process, we selected the scans with $\tau_{\mathrm{225}}<0.055\;(0.1)$ for 450~(850)~$\micron$ to ensure that all the data we are going to use were taken under good weather conditions. The $\tau_{\mathrm{225}}$ quantity here refers to the mean of ``WVMTAUST'' and ``WVMTAUEN'' (where ``ST'' and ``EN'' represent ``start'' and ``end,'' respectively) in the FITS header.\\

(3) To convert the picowatt (pW) units to millijansky per beam (mJy beam$^{-1}$), we applied the flux conversion factors (FCFs), which are usually in units of Jy~beam$^{-1}$~pW$^{-1}$, to our maps. We followed the guideline for Starlink version 2018A on the JCMT website.\footnote{\url{https://www.eaobservatory.org/jcmt/instrumentation/continuum/scuba-2/calibration/}} This guideline adopts the updated opacity relations at both 450~$\micron$ and 850~$\micron$ for the new atmospheric extinction corrections and the updated FCFs derived from the regular calibrator observations from mid-2011 to early 2021 published by \citet{mairsDecadeSCUBA2Comprehensive2021}. We also applied the optional corrections for decreases in the peak FCF in the evening and increases in the morning, as outlined in the guideline.
Besides the peak FCFs, we also corrected for the flux loss caused by the data reduction processes. This flux loss can be assessed by comparing the flux densities of the injected idealized point sources at the \texttt{makemap} step and the measured flux densities after going through all the reduction processes. We adopt a flux loss fraction of (5.1$\pm$0.3)\% for 450~$\micron$ and (10.90$\pm$0.02)\% for 850~$\micron$, as previously estimated in \citet{limSCUBA2UltraDeep2020}. We compensated the flux loss in the FCFs and then applied the compensated FCFs to the scans using the PIpeline for Combining and Analyzing Reduced Data \citep[PICARD,][]{jennessJCMTScienceArchive2008} with the \texttt{CALIBRATE\_SCUBA2\_DATA} recipe.\\

(4) To mosaic and co-add all the calibrated scans into a single map, we used the \texttt{MOSAIC\_JCMT\_IMAGES} recipe in PICARD with the default parameters. This recipe considers the weight of each pixel provided by \texttt{makemap} while combining the scans, to achieve the optimal signal-to-noise ratio.\\

(5) There exist isolated bad pixels that behave like compact sources after the map is convolved with a Gaussian kernel in Step (6). These probably come from spikes in the time stream data that escaped the filtering in Step (1). To prevent such bad pixels from becoming spurious detections, we removed them before Step (6) with median and standard-deviation ($\sigma$) filters. We identified isolated bad pixels as those exceeding the local median by $\pm4\sigma$. The local median and $\sigma$ were both measured within a box of $5\times5$ pixels, and the 4$\sigma$ threshold was determined by trial and error and eye inspection of the maps. Those identified as bad pixels were then replaced by the local median.\\

(6) To achieve optimal point-source detection, we applied a matched filter to our map with the \texttt{SCUBA2\_MATCHED\_FILTER} recipe in PICARD. This recipe consists of two steps: large-scale background removal and point-source enhancement. It first convolves our 450-$\micron$ (850-$\micron$) map with a 20$\arcsec$ (30$\arcsec$) FWHM Gaussian kernel to suppress the signal from point sources, leaving only the large-scale background in the map. It then subtracts this large-scale background map from the original map and convolves the map with a $7\farcs5$ (14\arcsec) FWHM Gaussian kernel, which is close to the instrumental beam size. This enhances the signal from point sources because sources with characteristic radii close to the size of the convolution kernel get the strongest signal boost after convolution \citep{stetsonDAOPHOTComputerProgram1987}, but comes at the cost of increasing blending. The resultant beam FWHM is $10\farcs3$ at 450 $\micron$ and $15\farcs2$ at 850 $\micron$ (Section \ref{psf}). This damages our ability to resolve close pairs in our source extraction, especially at 450 $\micron$. However, we note that the positional accuracy of source extraction for single sources still corresponds to that of the original diffraction resolution. This is verified with our counterpart identifications using high-resolution data from ALMA and VLA (Gao et al.\ in prep). Therefore in our subsequent discussion in this paper, ``FWHM'' means the original diffraction FWHM (i.e., before match-filtering) unless otherwise specified.

\subsection{Noise Estimation}\label{section_noise}
In reducing SCUBA-2 observations, three types of noise are of interest---instrumental noise ($\sigma_{\mathrm{i}}$), confusion noise ($\sigma_{\mathrm{c}}$), and total noise ($\sigma_{\mathrm{total}}$). The instrumental noise can be interpreted as the standard error of the mean of each pixel; if we keep increasing the integration time, this will become lower and lower. The instrumental noise comes naturally as an output map from \texttt{makemap}. We measured the instrumental noise on the matched-filter noise maps. The instrumental noise in the deepest parts of the maps (i.e., the innermost white dotted contours in Figure~\ref{fig:coverage_map}) is 0.59~mJy~beam$^{-1}$ at 450~$\micron$ and 0.09~mJy~beam$^{-1}$ at 850~$\micron$.

The total noise can be measured as the standard deviation of the source-masked area with a similar integration time. It consists of contributions from statistical uncertainty (i.e., $\sigma_{\mathrm{i}}$) and from unresolved crowded faint sources (also known as ``confusion noise,'' $\sigma_{\mathrm{c}}$). Therefore, the total noise can be expressed as \begin{equation}
    \sigma^{2}_{\mathrm{total}} = \sigma^{2}_{\mathrm{i}} + \sigma^{2}_{\mathrm{c}}.
\end{equation}
If the integration time is long enough, the instrumental noise becomes smaller and the confusion noise becomes dominant. Here, we estimate the confusion noise by comparing the measured total and instrumental noise levels.

\begin{figure}
\epsscale{2.2}
\plottwo{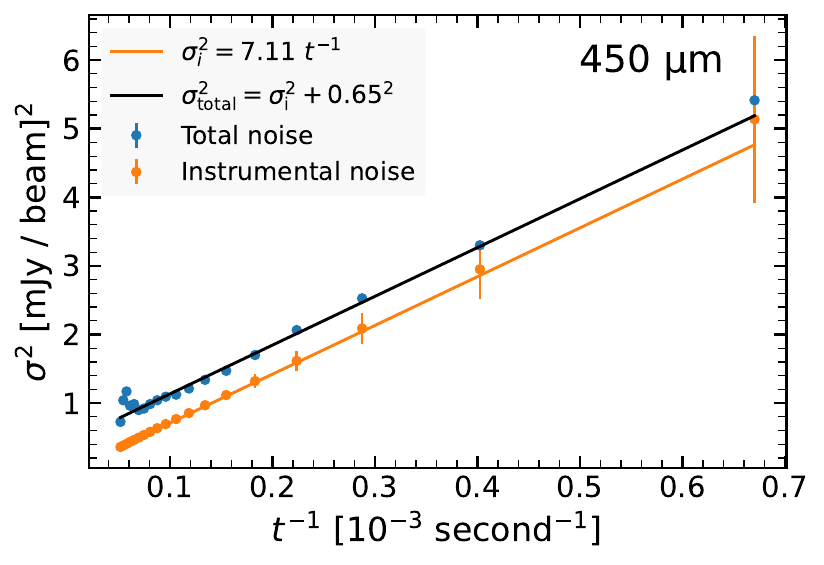}{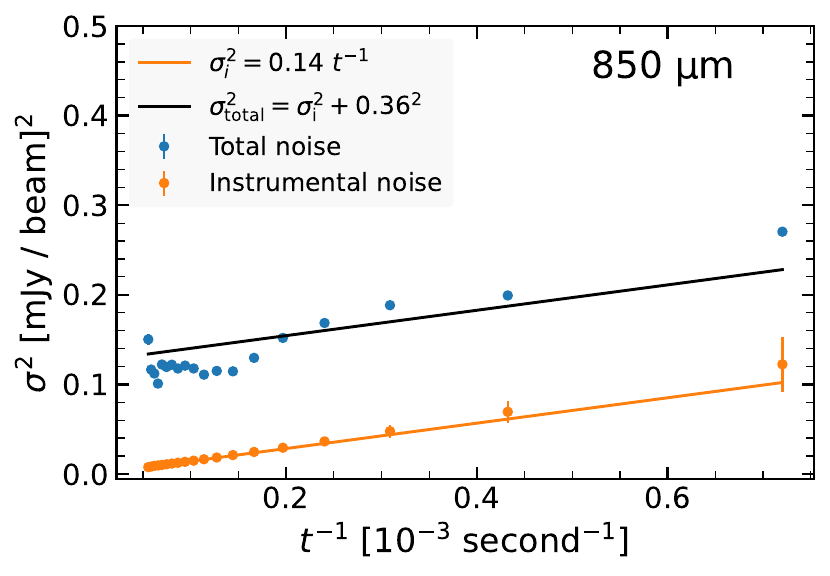}
\caption{Measurements of the confusion noise of STUDIES-COSMOS at 450~$\micron$ and 850~$\micron$. We cumulatively co-added the scans into 25 progressively deeper maps and used all the pixels to measure the instrumental noise and total noise binned by integration time. The best-fit values of the confusion noise are 0.65~mJy at 450~$\micron$ and 0.36~mJy at 850~$\micron$. It can be seen that the instrumental noise levels follow the $t^{-1}$ lines well in both wavebands. The total noise levels have larger deviations from the $t^{-1}$ lines, especially in deep regions and at 850~$\micron$, where confusion noise dominates. The error bars represent bootstrapped errors and are smaller than the symbols in most cases.}
\label{fig:confusion_noise}
\end{figure}

To perform the measurements, we divided our scans into 25 chunks of 40 scans each, mosaicked each chunk into a single map, and then cumulatively mosaicked the maps into 25 progressively deeper maps of the same field. To mask sources, we identified them by locating local maxima with S/N~$>3.5$ in the deepest map. We created a source mask in which each source is masked by a circle with a diameter of 3 times the beam FWHM, and applied the source mask to all of the 25 maps. We collected unmasked pixels from these maps, binned them based on their integration times, and measured $\sigma_{\mathrm{total}}$ within each bin. We then fit $\sigma^{2}_{\mathrm{i}}$ as a linear function of inverse exposure time with a fixed intercept of 0 at $t^{-1}=0$ (i.e., instrumental noise approaches zero as the integration time approaches infinity). We used the best-fit slope as a fixed parameter to find the best-fit intercept of $\sigma^{2}_{\mathrm{total}}$ (with bootstrapped uncertainties), and this intercept will be the square of the confusion noise ($\sigma^{2}_{\mathrm{c}}$). The concept of this procedure is illustrated in Figure \ref{fig:confusion_noise}. We found the best-fit 3.5$\sigma$-source-masked confusion noise to be $0.65\pm0.02$ and $0.36\pm0.01$~mJy~beam$^{-1}$ at 450 and 850~$\micron$, respectively. It is interesting to note that the points representing the 850 $\micron$ total noise in Figure \ref{fig:confusion_noise} do not imply comparable confusion noises in the deep and shallow regions. This may be  caused by variance in densities of faint confusing sources in different parts of the map.  We tested this hypothesis by splitting the shallow region into four quadrants, and we found a similar level of fluctuation in the measured confusion noises in the four quadrants.  The above fitted confusion noise of 0.36~mJy~beam$^{-1}$ sits between the deep and shallow regions, and should be sufficiently representative.

The above estimates of confusion noise are comparable to those of \citet{limSCUBA2UltraDeep2020}, but much lower than the 0.8~mJy~beam$^{-1}$ at 850~$\micron$ estimated by \citet{geachSCUBA2CosmologyLegacy2017}. We found that the higher value from \citet{geachSCUBA2CosmologyLegacy2017} may be due to contamination from bright sources that should have been masked during the estimation process. In the deepest regions of the maps, the instrumental noise levels are comparable to (450~$\micron$) and lower than (850~$\micron$) the confusion noise levels. This confirms that our maps at both 450 and 850~$\micron$ have reached the confusion limits.
In Appendix \ref{apd:confusion_limit}, we estimate the confusion limit by integrating the best-fit Schechter function (see Section \ref{subsec:corrected_number_counts}) until the rule-of-thumb source density criterion of one source per 30 beams is met. Although we have added the confusion noise to the instrumental noise map, we have still detected sources below the confusion limit. Sources below this limit should be used with caution (e.g., considering completeness and spurious probability).

\subsection{PSF modeling} \label{psf}

\begin{figure}
\epsscale{2.2}
\plottwo{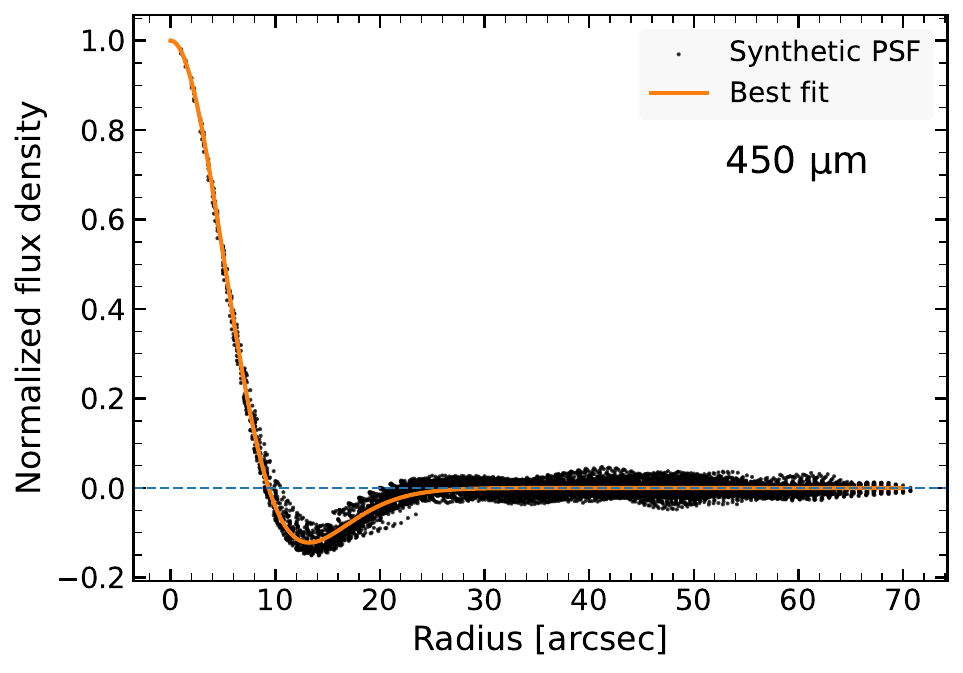}{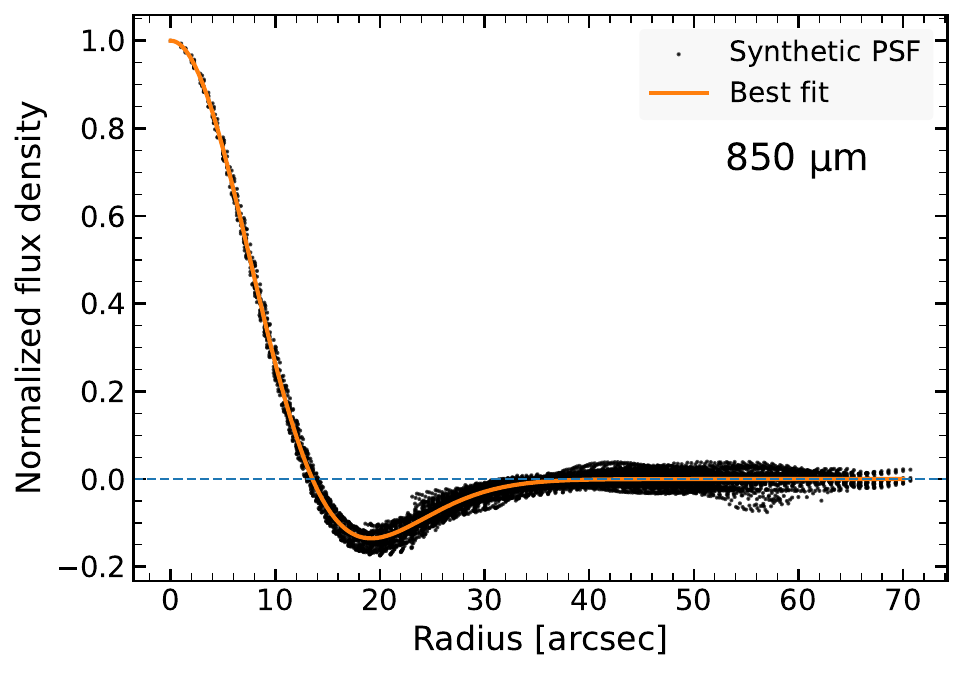}
\caption{Models of STUDIES-COSMOS PSFs at 450~$\micron$ and 850~$\micron$. The synthetic PSFs (black dots) are obtained by stacking high S/N ($>$7.9 at 450~$\micron$; $>$5 at 850~$\micron$) and isolated sources (separation $\gtrsim3\times$ FWHM) on the S/N images. The best-fit model (orange curves) consists of a double Gaussian function and shows a trough (a negative ring) caused by the large-scale background removal in the matched-filtering step.}
\label{fig:psf}
\end{figure}

To model the point spread function (PSF), for source extraction, we first created a synthetic PSF by stacking the S/N images of sources having no neighbor with S/N $>3$ within a radius of $3\times$FWHM (29\farcs4\ for 450~$\micron$ and 40$\arcsec$ for 850~$\micron$) to reduce the contamination from neighboring sources. The S/N cuts for the sources to be stacked are 7.9 (5.0) for 450 (850) $\micron$. We then set the peak intensity of the $101\arcsec \times 101\arcsec$ synthetic PSF to 1.0 and the total intensity to 0. The zero sum is a natural expectation from the matched-filtering processes in Step (6) of Section~\ref{data_reduction}. The profiles of the stacked PSFs are shown in Figure~\ref{fig:psf}. We note that observing the zero sum in the 2-D profile is not straightforward. However, in the actual 3-D PSF image, the trough covers a larger area due to its position at larger radii. Following previous work \citep[e.g.,][]{geachSCUBA2CosmologyLegacy2017,simpsonEastAsianObservatory2019}, we used a double Gaussian function to model the radial profile of the synthetic PSF,
\begin{equation}
    G(r) = A \exp{\left( -\frac{r^2}{2\sigma^2} \right)} - (A - 1) \exp{\left( -\frac{r^2}{2\sigma^2}\frac{A-1}{A} \right)},
\end{equation}
where $r$ is the radius. We found best-fit values of $A=1.84$ and $\sigma=5\farcs51$ for 450~$\micron$, and $A=246.60$ and $\sigma=9\farcs57$ for 850~$\micron$. The corresponding FWHMs of the fitted PSFs are 10\farcs3 and 15\farcs2 respectively.  In Section~\ref{source_extraction}, we use these PSF models for source extraction.

\subsection{Astrometry}

To assess the astrometric accuracy, we stacked the 450- and 850-$\micron$ S/N images based on the coordinates of 709 radio sources from the Very Large Array (VLA) 3-GHz catalog of \citet{smolcicVLACOSMOSGHzLarge2017}. The idea is that the majority of the radio sources should also be faint submillimeter sources.  Even if some radio sources have brighter and unrelated neighboring submillimeter sources, their effect should become negligible after stacking the 709 radio sources. Therefore, assuming that the radio coordinates are accurate, the centroids of the stacked SCUBA-2 images should indicate the astrometric offsets of the SCUBA-2 observations. The measured astrometric offsets in arcsec are $\Delta \mathrm{RA}=-0.10 \pm 0.12$ and $\Delta \mathrm{Dec}=0.17 \pm 0.13$ at 450~$\micron$ and $\Delta \mathrm{RA}=0.06 \pm 0.26$ and $\Delta \mathrm{Dec}=0.03 \pm 0.26$ at 850~$\micron$. The uncertainty was estimated using the bootstrap method, which involves repeated sampling with replacement. The results indicate that there is little astrometric offset in the SCUBA-2 data. Since these offsets are smaller than the pixel scale of one arcsec per pixel, no astrometric correction was applied to the maps.

\subsection{Source extraction} \label{source_extraction}

The algorithm we use to extract sources from our maps is similar to the CLEAN algorithm widely used in radio interferometric imaging. Our algorithm consists of three iterative steps: (1) finding the highest S/N peak with S/N $>$ 3.5; (2) recording the peak coordinate as the source coordinate and using this coordinate in the following steps if no previous record can be found within 4$\arcsec$ search radius for 450~$\micron$ (7$\arcsec$ for 850~$\micron$, both about half the beam FWHM), or using the previously recorded coordinate instead; and (3) subtracting 5\% of the peak-scaled PSF at that coordinate from the flux density map and recording the subtracted flux. The 5\% iterative approach aims to reveal nearby secondary and possibly tertiary peaks beyond the primary peak, given the blending due to poor angular resolution, by gradually subtracting 5\% of the flux  until the blended secondary (or tertiary) peak appears. Once all peaks above 3.5$\sigma$ were subtracted, the iterative procedure stops. For each source, we obtained the raw flux density by reading the residual flux density in the residual map at the recorded coordinate and summing it with the total subtracted flux density. This raw flux density will be further corrected for the flux-boosting effect using the Monte Carlo simulations described in Section \ref{simulations}. At 450~$\micron$ we detect 360 $>4\sigma$ sources and 479 $>3.5\sigma$ sources, with expected false detection rates of 11.5\% and 19.6\%, respectively. At 850~$\micron$ we detect 237 $>4\sigma$ sources and 314 $>3.5\sigma$ sources, with expected false detection rates of 10.5\% and 16.6\%, respectively. Within 7-arcsec search radii (i.e., half the 850-$\micron$ FWHM), we found that 151 (63.7\%) of the 237 850-$\micron$ sources have at least one associated $>4\sigma$ 450-$\micron$ source. The extracted sources are listed in Table \ref{tab:450_catalog} (450~$\micron$) and Table \ref{tab:850_catalog} (850~$\micron$). Note that the $\sigma$ value here represents the instrumental noise at 450~$\micron$ and the total noise at 850~$\micron$. This choice is because the total noise at 850~$\micron$ is entirely dominated by the confusion noise rather than instrumental noise.

\begin{deluxetable*}{ccccccccc}
\tablecaption{\label{tab:450_catalog}STUDIES-COSMOS 450-$\micron$ source catalog}
\tablehead{\colhead{ID} & \colhead{RA} & \colhead{Dec} & \colhead{S/N} & \colhead{$S_{\mathrm{obs}}$} & \colhead{$S_{\mathrm{corr}}$} & \colhead{Comp.} & \colhead{Spur.} \\ 
\colhead{} & \colhead{(J2000)} & \colhead{(J2000)} & \colhead{} & \colhead{(mJy)} & \colhead{(mJy)} & \colhead{(\%)} & \colhead{(\%)}} 

\startdata
STUDIES-COSMOS-450-001 & 10 00 33.37 & +02 26 00.00 & 44.9 & 27.6 $\pm$ 0.6 & 27.2 $\pm$ 1.6 & 100.0 &   0.0 \\
STUDIES-COSMOS-450-002 & 10 00 39.24 & +02 22 21.00 & 35.6 & 28.9 $\pm$ 0.8 & 28.5 $\pm$ 1.7 &  99.9 &   0.0 \\
STUDIES-COSMOS-450-003 & 10 00 23.69 & +02 21 56.00 & 29.1 & 22.4 $\pm$ 0.8 & 22.0 $\pm$ 1.6 &  99.9 &   0.0 \\
STUDIES-COSMOS-450-004 & 10 00 34.37 & +02 21 22.00 & 26.6 & 20.7 $\pm$ 0.8 & 20.3 $\pm$ 1.7 & 100.0 &   0.2 \\
STUDIES-COSMOS-450-005 & 10 00 28.76 & +02 32 02.00 & 25.5 & 30.8 $\pm$ 1.2 & 30.1 $\pm$ 2.1 & 100.0 &   0.2 \\
STUDIES-COSMOS-450-006 & 10 00 25.30 & +02 18 47.00 & 23.1 & 24.1 $\pm$ 1.1 & 23.5 $\pm$ 2.0 & 100.0 &   0.4 \\
STUDIES-COSMOS-450-007 & 10 00 16.69 & +02 26 38.00 & 22.5 & 19.4 $\pm$ 0.9 & 18.9 $\pm$ 1.7 &  99.9 &   0.3 \\
STUDIES-COSMOS-450-008 & 10 00 28.56 & +02 27 25.00 & 22.5 & 13.8 $\pm$ 0.6 & 13.4 $\pm$ 1.5 &  99.7 &   0.4 \\
STUDIES-COSMOS-450-009 & 10 00 25.50 & +02 25 44.00 & 21.4 & 14.1 $\pm$ 0.7 & 13.8 $\pm$ 1.5 &  99.9 &   0.6 \\
STUDIES-COSMOS-450-010 & 10 00 08.21 & +02 26 12.99 & 21.0 & 26.0 $\pm$ 1.2 & 25.5 $\pm$ 2.1 & 100.0 &   0.0 \\
\enddata

\tablecomments{$S_{\mathrm{obs}}$ gives the observed flux density and the instrumental noise. $S_{\mathrm{corr}}$ gives the de-boosted flux density and the total noise (instrumental, confusion, and de-boosting). ``Comp.'' represents the completeness of the source. ``Spur.'' is the spurious probability of the source.}
(This table is available in its entirety in the machine-readable format.)
\end{deluxetable*}

\begin{deluxetable*}{ccccccccc}
\tablecaption{\label{tab:850_catalog}STUDIES-COSMOS 850-$\micron$ source catalog}
\tablehead{\colhead{ID} & \colhead{RA} & \colhead{Dec} & \colhead{S/N} & \colhead{$S_{\mathrm{obs}}$} & \colhead{$S_{\mathrm{corr}}$} & \colhead{Comp.} & \colhead{Spur.} \\ 
\colhead{} & \colhead{(J2000)} & \colhead{(J2000)} & \colhead{} & \colhead{(mJy)} & \colhead{(mJy)} & \colhead{(\%)} & \colhead{(\%)}} 

\startdata
STUDIES-COSMOS-850-001 & 10 00 08.15 & +02 26 11.99 & 43.5 & 17.1 $\pm$ 0.4 & 17.0 $\pm$ 0.7 &  99.9 &   0.0 \\
STUDIES-COSMOS-850-002 & 10 00 15.62 & +02 15 49.00 & 31.2 & 14.4 $\pm$ 0.5 & 14.3 $\pm$ 0.8 &  99.9 &   0.0 \\
STUDIES-COSMOS-850-003 & 10 00 19.82 & +02 32 04.00 & 28.9 & 11.8 $\pm$ 0.4 & 11.7 $\pm$ 0.7 &  99.9 &   0.0 \\
STUDIES-COSMOS-850-004 & 10 00 28.76 & +02 32 04.00 & 27.8 & 11.2 $\pm$ 0.4 & 11.1 $\pm$ 0.7 &  99.9 &   0.0 \\
STUDIES-COSMOS-850-005 & 10 00 33.44 & +02 25 59.00 & 25.1 &  9.3 $\pm$ 0.4 &  9.0 $\pm$ 0.6 &  99.9 &   0.2 \\
STUDIES-COSMOS-850-006 & 10 00 23.96 & +02 17 51.00 & 24.6 &  9.8 $\pm$ 0.4 &  9.8 $\pm$ 0.7 &  99.8 &   0.1 \\
STUDIES-COSMOS-850-007 & 10 00 23.63 & +02 21 55.00 & 21.5 &  8.1 $\pm$ 0.4 &  8.0 $\pm$ 0.7 &  99.7 &   0.3 \\
STUDIES-COSMOS-850-008 & 10 00 34.37 & +02 21 22.00 & 19.8 &  7.5 $\pm$ 0.4 &  7.4 $\pm$ 0.6 &  99.6 &   0.3 \\
STUDIES-COSMOS-850-009 & 10 00 49.92 & +02 22 58.99 & 19.4 &  7.7 $\pm$ 0.4 &  7.6 $\pm$ 0.7 &  99.8 &   0.2 \\
STUDIES-COSMOS-850-010 & 10 00 25.23 & +02 26 07.00 & 20.3 &  7.5 $\pm$ 0.4 &  7.5 $\pm$ 0.6 &  99.6 &   0.3 \\
\enddata

\tablecomments{$S_{\mathrm{obs}}$ gives the observed flux density and the total noise consisting of the instrumental noise and the confusion noise. This is different from the one in Table \ref{tab:450_catalog} because the instrumental noise of the 850-$\micron$ map is well below the confusion noise. $S_{\mathrm{corr}}$ gives the de-boosted flux density and the combined noise of instrumental noise, confusion noise, and de-boosting uncertainty. ``Comp.'' represents the completeness of the source. ``Spur.'' represents the spurious probability of the source.}
(This table is available in its entirety in the machine-readable format.)
\end{deluxetable*}

\section{Number Counts} \label{sec:number_counts}
\subsection{Raw number counts} \label{sec:raw_counts}

\begin{figure}
\epsscale{2.3}
\plottwo{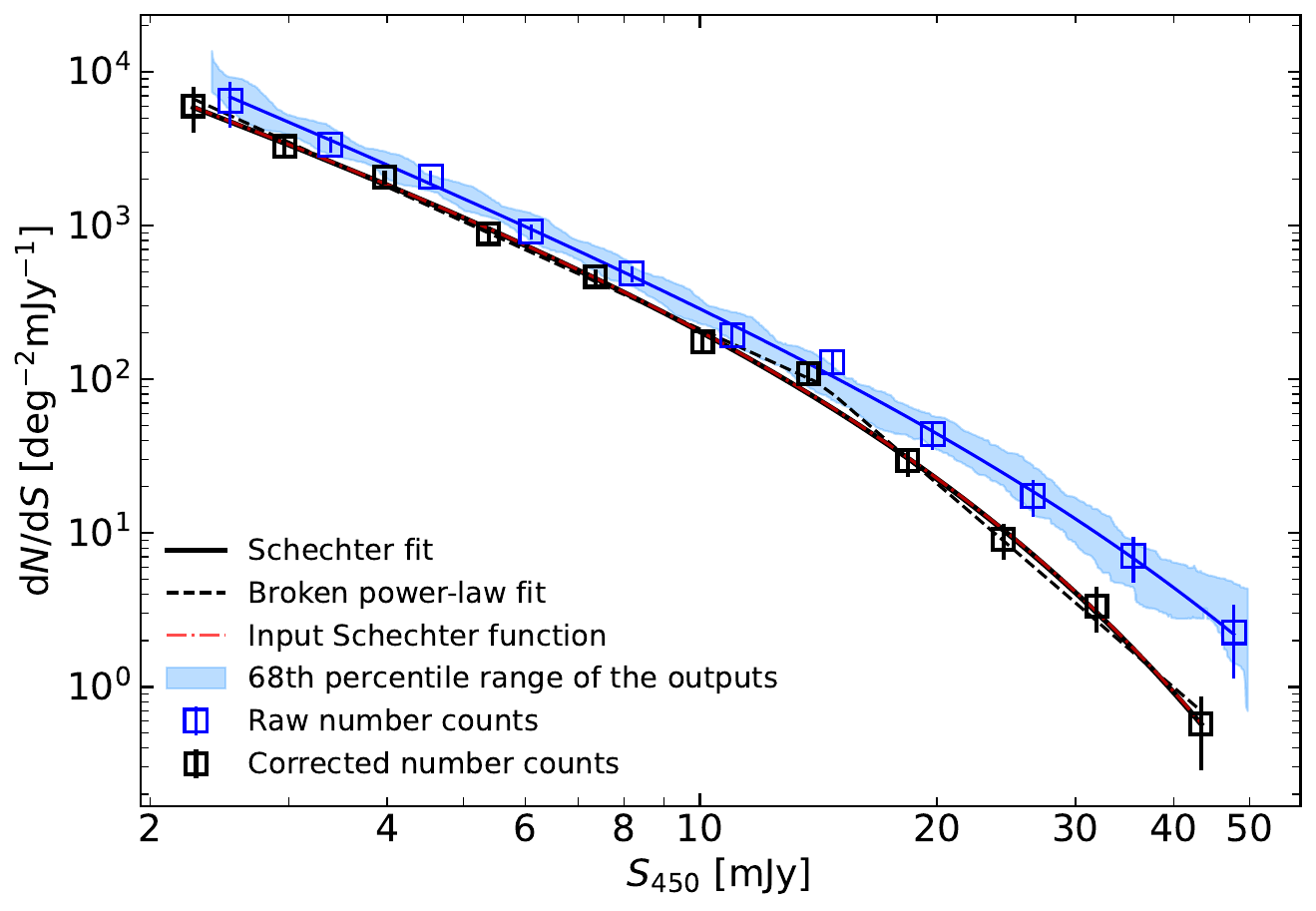}{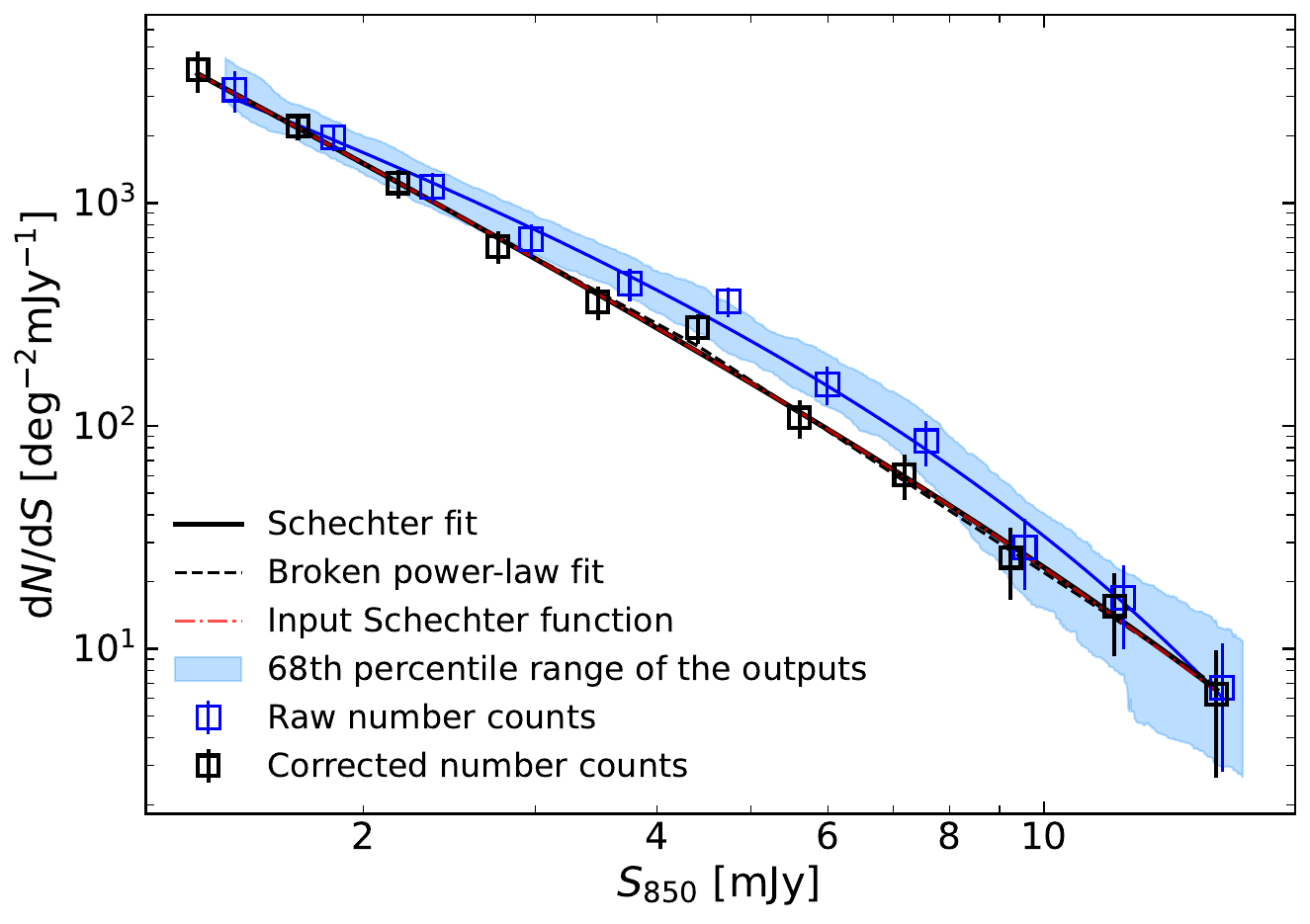}
\caption{Differential number counts at 450~$\micron$ and 850~$\micron$. The blue and black symbols show the raw counts (see Section~\ref{sec:raw_counts}) and the corrected counts (see Section~\ref{subsec:corrected_number_counts}), respectively. The blue and black solid curves are the best-fit Schechter functions to the raw and corrected counts, respectively. The black dashed curves are the best-fit broken power-law functions to the corrected counts. The red dash-dotted curves show the input Schechter function used in the last iteration of the simulations. The convergence of our iterative procedures can be verified by the fact that the red and black solid curves are indistinguishable. The blue shaded region shows the 68th percentile range of the output counts of the 400 simulated maps generated in the last iteration of the simulations, which are generally consistent with the Poissonian error bars of the raw counts.}
\label{fig:raw_corr}
\end{figure}

We derived the 450-$\micron$ and 850-$\micron$ raw number counts from the 3.5$\sigma$ source catalogs extracted in Section \ref{source_extraction} along with the noise maps. For each flux density bin, we calculated the number density of sources in that flux density interval by summing up $\frac{1}{A_{\rm e}(S)}$ of each source, where $A_{\rm e}(S)$ is the effective area for a source of flux density $S$ where it can be detected at $>3.5\sigma$. The error based on Poisson statistics with error propagation can be calculated by $\sqrt{\Sigma (\frac{\sqrt{1}}{A_{\rm e}(S)})^{2}}$ for each flux density bin. Practically, $A_{\rm e}(S)$ can be estimated by counting the number of pixels with noise less than $S/3.5$ on the noise map and then multiplying that number by the area per pixel. After calculating the number density of each bin, we divided the number densities and their errors by their flux density intervals $dS$ to obtain the differential raw number counts. The raw counts shown in Figure \ref{fig:raw_corr} can be approximately described by a Schechter function,
\begin{equation}
\label{eq:schechter}
\frac{dN}{dS} = \left(\frac{N_{0}}{S_{0}}\right) \left(\frac{S}{S_{0}}\right)^{\alpha} \exp{\left(-\frac{S}{S_{0}}\right)}.
\end{equation}
We use the best-fit functions as initial inputs in the simulations for estimating the intrinsic number counts in Section \ref{simulations}.

\begin{deluxetable*}{ccccccc}
\tablecaption{\label{tab:450_counts}450-$\micron$ number counts}
\tablehead{\colhead{$N$} & \colhead{$S_{\mathrm{obs}}$} & \colhead{Raw $dN/dS$} & \colhead{$S_{\mathrm{corr}}$} & \colhead{Corrected $dN/dS$} & \colhead{$S_{\mathrm{corr}}$} & \colhead{Corrected $N(>S)$} \\ 
\colhead{} & \colhead{(mJy)} & \colhead{(deg$^{-2}$ mJy$^{-1}$)} & \colhead{(mJy)} & \colhead{(deg$^{-2}$ mJy$^{-1}$)} & \colhead{(mJy)} & \colhead{(deg$^{-2}$)} } 

\startdata
20 & 2.53 & 6519 $\pm$ 2155 & 2.27 & 6010 $\pm$ 1987 & 2.06 & 21353 $\pm$ 2965 \\
73 & 3.39 & 3369 $\pm$ 408 & 2.96 & 3269 $\pm$ 396 & 2.72 & 11602 $\pm$ 644 \\
104 & 4.55 & 2089 $\pm$ 206 & 3.97 & 2065 $\pm$ 204 & 3.61 & 7784 $\pm$ 436 \\
82 & 6.10 & 913 $\pm$ 101 & 5.38 & 889 $\pm$ 99 & 4.78 & 4396 $\pm$ 280 \\
69 & 8.18 & 488 $\pm$ 59 & 7.36 & 467 $\pm$ 57 & 6.32 & 2727 $\pm$ 206 \\
41 & 11.0 & 193 $\pm$ 31 & 10.1 & 175 $\pm$ 28 & 8.38 & 1516 $\pm$ 142 \\
41 & 14.7 & 129 $\pm$ 21 & 13.7 & 109 $\pm$ 17 & 11.1 & 961 $\pm$ 108 \\
22 & 19.8 & 44.3 $\pm$ 9.5 & 18.4 & 29.7 $\pm$ 6.4 & 14.7 & 437 $\pm$ 67 \\
14 & 26.5 & 17.4 $\pm$ 4.7 & 24.3 & 9.10 $\pm$ 2.44 & 19.5 & 196 $\pm$ 42 \\
9 & 35.6 & 7.11 $\pm$ 2.37 & 31.9 & 3.36 $\pm$ 1.12 & 25.8 & 85.4 $\pm$ 26.3 \\
4 & 47.7 & 2.27 $\pm$ 1.14 & 43.3 & 0.57 $\pm$ 0.29 & 34.1 & 15.8 $\pm$ 11.2 \\
\enddata

\tablecomments{$S_{\mathrm{obs}}$ is the observed flux density. $S_{\mathrm{corr}}$ is the de-boosted flux density. For $S_{\mathrm{obs}}$, the bin widths are identical on the logarithmic scale, and the logarithmic centers are used as the bin centers.}
\end{deluxetable*}

\begin{deluxetable*}{ccccccc}
\tablecaption{\label{tab:850_counts}850-$\micron$ number counts}
\tablehead{\colhead{$N$} & \colhead{$S_{\mathrm{obs}}$} & \colhead{Raw $dN/dS$} & \colhead{$S_{\mathrm{corr}}$} & \colhead{Corrected $dN/dS$} & \colhead{$S_{\mathrm{corr}}$} & \colhead{Corrected $N(>S)$} \\ 
\colhead{} & \colhead{(mJy)} & \colhead{(deg$^{-2}$ mJy$^{-1}$)} & \colhead{(mJy)} & \colhead{(deg$^{-2}$ mJy$^{-1}$)} & \colhead{(mJy)} & \colhead{(deg$^{-2}$)} } 

\startdata
38 & 1.47 & 3231 $\pm$ 689 & 1.35 & 3957 $\pm$ 844 & 1.27 & 6429 $\pm$ 699 \\
53 & 1.86 & 1977 $\pm$ 272 & 1.71 & 2208 $\pm$ 304 & 1.61 & 3251 $\pm$ 221 \\
46 & 2.35 & 1184 $\pm$ 175 & 2.17 & 1231 $\pm$ 182 & 2.03 & 2153 $\pm$ 158 \\
37 & 2.97 & 686 $\pm$ 113 & 2.75 & 640 $\pm$ 106 & 2.57 & 1542 $\pm$ 126 \\
35 & 3.75 & 435 $\pm$ 74 & 3.48 & 361 $\pm$ 61 & 3.26 & 956 $\pm$ 91 \\
43 & 4.74 & 362 $\pm$ 56 & 4.41 & 276 $\pm$ 42 & 4.13 & 647 $\pm$ 72 \\
26 & 5.99 & 154 $\pm$ 31 & 5.61 & 109 $\pm$ 22 & 5.23 & 350 $\pm$ 52 \\
19 & 7.56 & 85.5 $\pm$ 19.7 & 7.19 & 60.6 $\pm$ 13.9 & 6.62 & 185 $\pm$ 38 \\
8 & 9.55 & 28.5 $\pm$ 10.1 & 9.24 & 25.7 $\pm$ 9.1 & 8.39 & 93.1 $\pm$ 27.0 \\
6 & 12.1 & 16.9 $\pm$ 6.9 & 11.8 & 15.6 $\pm$ 6.4 & 10.6 & 70.0 $\pm$ 23.4 \\
3 & 15.2 & 6.69 $\pm$ 3.87 & 15.0 & 6.28 $\pm$ 3.63 & 13.4 & 22.8 $\pm$ 13.2 \\
\enddata

\tablecomments{$S_{\mathrm{obs}}$ is the observed flux density. $S_{\mathrm{corr}}$ is the de-boosted flux density. For $S_{\mathrm{obs}}$, the bin widths are identical on the logarithmic scale, and the logarithmic centers are used as the bin centers.}

\end{deluxetable*}

\subsection{Simulations and final catalogs} \label{simulations}

The raw number counts above are biased by several observational effects: flux boosting caused by Eddington-type biases and faint undetected sources; detection incompleteness; presence of spurious sources; and source blending. To recover the intrinsic number counts, we carried out Monte Carlo simulations. The simulation process involves randomly injecting sources into a ``true-noise map'' (see below) with a flux density distribution following a certain function (e.g., the best-fit function of the observed raw number counts). The output of this simulation contains the biases caused by the observational effects mentioned. By iteratively comparing the output counts with the observed raw number counts, we adjust the input for each iteration until the output converges with the observed raw number counts. This convergence confirms that the final input is a close approximation of the intrinsic number counts.

One of the important elements in the simulations is the true-noise map into which we injected the simulated sources. We created the true-noise map using the jackknife method first introduced by \citet{cowieFaintSubmillimeterCounts2002} for SCUBA images. We sorted the calibrated scans made in Section \ref{data_reduction} by the dates of the observations and divided the scans into odd and even parts. The two parts of the scans went through the same procedures of Steps (4) to (6) in Section \ref{data_reduction} to form two maps. The way we separate the scans ensures that the two maps have similar weather-condition distributions and area coverages. We subtracted one map from the other and then multiplied the subtracted map by $\sqrt{t_{1}t_{2}}/(t_{1}+t_{2})$ on a pixel-to-pixel basis to scale down the rms noise level to match that of the final mosaicked map based on the relation of $\mathrm{rms}\sim t^{-1/2}$, where $t$ is the weighted exposure time. This operation effectively removes all sources in the map, including undetected faint sources, and provides a source-free true-noise map that can be used for simulations.

In practice, an iteration in the simulations involves the following six steps.
(1) Generation of input source catalogs: Using the best-fit Schechter function of the observed raw number counts as the initial input, we generate 400 catalogs with sources of flux densities ranging from 0.1 to 100~mJy, divided into 200 logarithmic bins, with more than 20,000 sources in each catalog. Poisson noise is added to the number of sources in each bin.

(2) Placement of sources: To create simulated maps, we randomly place sources from the 400 catalogs onto 400 noise-free maps. These source-only maps are then convolved with the PSF model from Section \ref{psf} to generate simulated source maps. We add the true-noise map to 200 of the simulated maps and the inverted true-noise map to the other 200 simulated maps.

(3) Source extraction: Using the same algorithm and parameters as in Section \ref{source_extraction}, sources are extracted from the 400 simulated maps to obtain 400 output source catalogs.

(4) Deriving output counts: The output number counts are derived from the combined output catalog, using the same flux density bins as the observed raw counts.

(5) Correction factor calculation: The goal is to converge the output and raw counts. The ratio between the two counts is calculated as the correction factor for each bin, which can be expressed as  $$C_{i}=\left(\frac{dN}{dS}\right)_{i,\rm{raw}}\bigg/\left(\frac{dN}{dS}\right)_{i,\rm{output}}.$$

(6) Adjusting input: The correction factors are used to adjust the input counts given by the input Schechter function of the current iteration. A Schechter function is fit to the adjusted input counts, and the best-fit function is used as the input for the next iteration.

Note that the simulations were conducted separately for the 450-$\micron$ and 850-$\micron$ maps. In our previous work \citep{wangSCUBA2UltraDeep2017}, we found that the simulation output and the observed raw counts typically converged after three iterations. Here, we ran ten iterations to ensure good convergence both between the input and the corrected output, and between the output and the observed raw counts. At 450~$\micron$, the results converge after three iterations at $\leq$10~mJy but seven iterations at $>$10~mJy possibly due to the larger uncertainties in the observed bright-end number counts.

\begin{figure*}
\epsscale{1.15}
\plotone{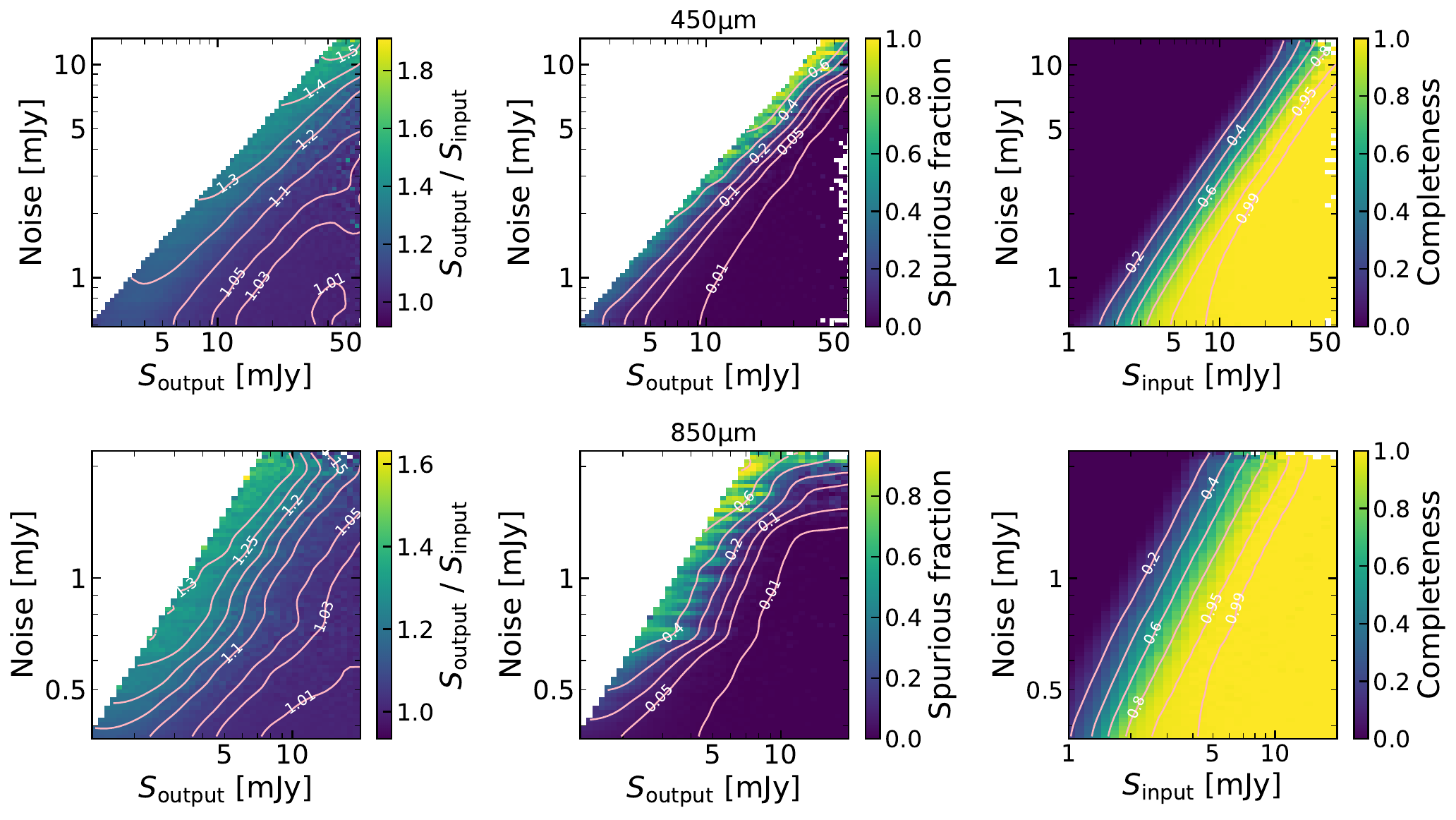}
\caption{Visualizations of different bias factors estimated from the Monte Carlo simulations as functions of flux density and local noise. The first column shows the flux-boosting factor (i.e., the ratio of output to input flux density). The second column shows the spurious fraction (the fraction of output sources without input counterparts). The third column shows the completeness (the fraction of input sources that are recovered in the output). The noise here refers to the instrumental noise at 450~$\micron$ and the total noise (i.e., including the confusion noise) at 850~$\micron$.}
\label{fig:factors}
\end{figure*}

The simulations allow us to estimate the flux-boosting factor, the spurious probability, and the completeness for each source based on its flux density and local noise level. To do this, we used the input Schechter function from the last iteration to generate more simulated maps. We generated 5,000 simulated maps using the positive true-noise map and another 5,000 maps using the negative true-noise map. After source extraction, we cross-matched the 10,000 input and output catalog pairs using the following criteria: (1) a search radius of 1/2 beam FWHM; (2) $S_{\mathrm{output}}/S_{\mathrm{input}}\leq 2$; and (3) taking the brightest source within the search radius as the counterpart. The flux-boosting factor is then estimated as $S_{\mathrm{output}} / S_{\mathrm{input}}$. The spurious probability is estimated as the fraction of output sources without input counterparts. The completeness is estimated as the fraction of input sources with output counterparts. Furthermore, the large number of 10,000 input and output catalog pairs allows us to compute these factors as functions of flux density and local noise. We show visualizations of each factor in Figure \ref{fig:factors}. The noise in the figure refers to the local instrumental noise at 450~$\micron$ and the total noise (including instrumental and confusion noise) at 850~$\micron$; this is because the instrumental noise is the dominant term at 450~$\micron$, becoming comparable to the confusion noise only at the center of the map. We estimated the flux-boosting factor, spurious probability, and completeness of each source by using linear interpolation to find the values corresponding to the flux density and local noise of each source. These values were then used to compile the final catalogs, which are presented in Table \ref{tab:450_catalog} (450~$\micron$) and Table \ref{tab:850_catalog} (850~$\micron$).

\subsection{Corrected number counts} \label{subsec:corrected_number_counts}

To correct the raw number counts, we applied corrections along both the ordinate ($S$) and the abscissa ($dN/dS$). We used the flux-boosting factor as a function of the output flux density to convert the observed flux density of the raw counts $S_{\mathrm{obs}}$ to the corrected flux density $S_{\mathrm{corr}}$. And we calculated the ratio of the output counts $C_{\mathrm{sim,output}}(S_{\mathrm{obs}})$ to the input counts $C_{\mathrm{sim,input}}(S_{\mathrm{corr}})$ to correct the counts. In this framework, the corrected counts are expressed as \begin{equation}
    C_{\mathrm{corr}}(S_{\mathrm{corr}}) = C_{\mathrm{raw}}(S_{\mathrm{obs}}) \times \frac{C_{\mathrm{sim,input}}(S_{\mathrm{corr}})}{C_{\mathrm{sim,output}}(S_{\mathrm{obs}})}.
\end{equation}
We present the corrected counts in Figure~\ref{fig:raw_corr} as black symbols. We list the raw and corrected differential number counts at 450 and 850~$\micron$ in Tables~\ref{tab:450_counts} and \ref{tab:850_counts}, respectively. We also list the cumulative number counts constructed using the final catalogs. We fit a Schechter function (Equation \ref{eq:schechter}) and a broken power law, \begin{equation}
\label{eq:bplaw}
\frac{dN}{dS} = 
    \begin{cases}
        N_{0}\left(\frac{S}{S_{0}}\right)^\alpha, & S \leq S_{0}, \\
        N_{0}\left(\frac{S}{S_{0}}\right)^\beta, & S > S_{0},
    \end{cases}
\end{equation}
to the corrected differential number counts at both 450 and 850~$\micron$. The best-fit parameters are listed in Tables \ref{tab:param_450} and \ref{tab:param_850}.

\begin{deluxetable}{ccc}
\tablecaption{\label{tab:param_450}Parameterizations for the corrected differential counts at 450~$\micron$}
\tablehead{\colhead{Parameter} & \colhead{Schechter fit} & \colhead{Broken power-law fit} \\ 
\colhead{} & \colhead{(Equation (\ref{eq:schechter}))} & \colhead{(Equation (\ref{eq:bplaw}))} } 

\startdata
$N_{0}$ & 5484$\pm$1658 [deg$^{-2}$] & 92.7$\pm$30.0 [deg$^{-2}$ mJy$^{-1}$] \\
$S_{0}$ & 10.1$\pm$1.5 [mJy] & 14.3$\pm$1.8 [mJy] \\
$\alpha$ & $-$1.75$\pm$0.16 & $-$2.33$\pm$0.09 \\
$\beta$ & \nodata & $-$4.41$\pm$0.41 \\
$\chi^2$ & 5.53 & 4.85\\
$\chi_{\nu}^2$ & 0.69 & 0.69 \\
\enddata

\end{deluxetable}

\begin{deluxetable}{ccc}
\tablecaption{\label{tab:param_850}Parameterizations for the corrected differential counts at 850~$\micron$}
\tablehead{\colhead{Parameter} & \colhead{Schechter fit} & \colhead{Broken power-law fit} \\ 
\colhead{} & \colhead{(Equation (\ref{eq:schechter}))} & \colhead{(Equation (\ref{eq:bplaw}))} } 

\startdata
$N_{0}$ & 290$\pm$350 [deg$^{-2}$] & 229$\pm$186 [deg$^{-2}$ mJy$^{-1}$] \\
$S_{0}$ & 14.5$\pm$9.0 [mJy] & 4.41$\pm$1.45 [mJy] \\
$\alpha$ & $-$2.25$\pm$0.19 & $-$2.37$\pm$0.11 \\
$\beta$ & \nodata & $-$2.86$\pm$0.26 \\
$\chi^2$ & 2.97 & 2.41\\
$\chi_{\nu}^2$ & 0.37 & 0.34 \\
\enddata

\end{deluxetable}

\section{Discussion} \label{sec:discussion}
\subsection{Comparison with Other Counts} \label{subsec:count_comparison}

Number counts, which are derived solely from images without any additional information like redshifts, provide a measure of the density of sources versus their flux density. At submillimeter to millimeter wavelengths, these number counts can be used to effectively constrain models of galaxy evolution \citep[e.g.,][]{baughCanFaintSubmillimetre2005,valianteBACKWARDEVOLUTIONMODEL2009,betherminUNIFIEDEMPIRICALMODEL2012,haywardSubmillimetreGalaxiesHierarchical2013,cowleySimulatedObservationsSubmillimetre2015,laceyUnifiedMultiwavelengthModel2016,betherminImpactClusteringAngular2017,cowleyEvolutionUVtommExtragalactic2019,lagosPhysicalPropertiesEvolution2020,poppingALMASpectroscopicSurvey2020,lovellReproducingSubmillimetreGalaxy2021}. Furthermore, by integrating the differential number counts, we can estimate the contributions of resolved SMGs to the EBL.
In order to gain a better understanding of the variance of the counts among different fields, and to better observe the behavior of the counts at different flux density ranges, we compare our counts with those in the literature. In this discussion, we will focus specifically on the 450-$\micron$ counts, as the 850-$\micron$ counts have already been well studied \citep[e.g.,][]{weissLargeApexBolometer2009,chenFaintSubmillimeterGalaxy2013,chenResolvingCosmicFarinfrared2013,hsuHawaiiSCUBA2Lensing2016,zavalaSCUBA2CosmologyLegacy2017,geachSCUBA2CosmologyLegacy2017,stachALMASurveySCUBA22018,simpsonEastAsianObservatory2019,betherminALPINEALMACIISurvey2020,shimNEPSC2NorthEcliptic2020,simpsonALMASurveyBrightest2020,chenALMACALIXMultiband2023}.

\subsubsection{Compilation of the Counts}
 
There have been multiple studies of the 450-$\micron$ number counts in the COSMOS field \citep{caseyCharacterizationSCUBA24502013,geachSCUBA2CosmologyLegacy2013,wangSCUBA2UltraDeep2017}. Our map combines data from all of these studies, as well as adding additional integration time to reach the confusion limit. In addition to the COSMOS field, we have also included the 450-$\micron$ counts of the EGS field from \citet{zavalaSCUBA2CosmologyLegacy2017}, the CDF-N and CDF-S fields from \citet{bargerSubmillimeterPerspectiveGOODS2022a}, the STUDIES-SXDS field, and the combined multi-field counts from \citet{chenFaintSubmillimeterGalaxy2013,chenResolvingCosmicFarinfrared2013} and \citet{hsuHawaiiSCUBA2Lensing2016}. The results of these fields will be discussed in more detail later. To make fair comparisons, we re-calibrated the data and re-constructed the number counts of the EGS field using the same method as in this study. However, we did not modify the data from \citet{chenResolvingCosmicFarinfrared2013} and \citet{hsuHawaiiSCUBA2Lensing2016} because working with lensing-cluster fields is outside the scope of our reduction pipeline. The combined multi-field counts at 450~$\micron$ from \citet{chenResolvingCosmicFarinfrared2013} consist of two lensing-cluster fields and one blank field (COSMOS data from \citealt{caseyCharacterizationSCUBA24502013}), while the counts from \citet{hsuHawaiiSCUBA2Lensing2016} include four lensing-cluster fields and the same blank field. The work by \citeauthor{hsuHawaiiSCUBA2Lensing2016} is an updated version of the work of \citeauthor{chenResolvingCosmicFarinfrared2013}, with enhanced sensitivity and two additional fields.

In addition to published 450-$\micron$ surveys, STUDIES includes a second pointing in the SXDS field, with observations still ongoing. We processed the STUDIES-SXDS data collected until June 2022, and the archival S2CLS 450-$\micron$ ``CV Daisy'' data for the Ultra Deep Survey (UDS) field, using the same method as for our STUDIES-COSMOS data. The STUDIES-SXDS map covers 130~arcmin$^2$ with an rms sensitivity of about 1~mJy at the center. The corrected 450-$\micron$ counts for STUDIES-SXDS can be found in Table \ref{tab:450_counts_sxds}. Further analysis of this field will be presented in a future paper.

\begin{deluxetable}{cc}
\tablecaption{\label{tab:450_counts_sxds}450-$\micron$ number counts in the STUDIES-SXDS Field}
\tablehead{\colhead{$S_{\mathrm{corr}}$} & \colhead{Corrected $dN/dS$} \\ 
\colhead{(mJy)} & \colhead{(deg$^{-2}$ mJy$^{-1}$)}} 

\startdata
4.10 & 2149 $\pm$ 518 \\
5.22 & 1265 $\pm$ 191 \\
6.82 & 509  $\pm$ 90 \\
9.01 & 417  $\pm$ 67 \\
12.0 & 162  $\pm$ 35 \\
15.8 & 56.2 $\pm$ 18.7 \\
20.7 & 45.1 $\pm$ 15.0
\enddata
\tablecomments{$S_{\mathrm{corr}}$ is the de-boosted flux density.}
\end{deluxetable}

\begin{figure*}[htbp]
\epsscale{1.17}
\plotone{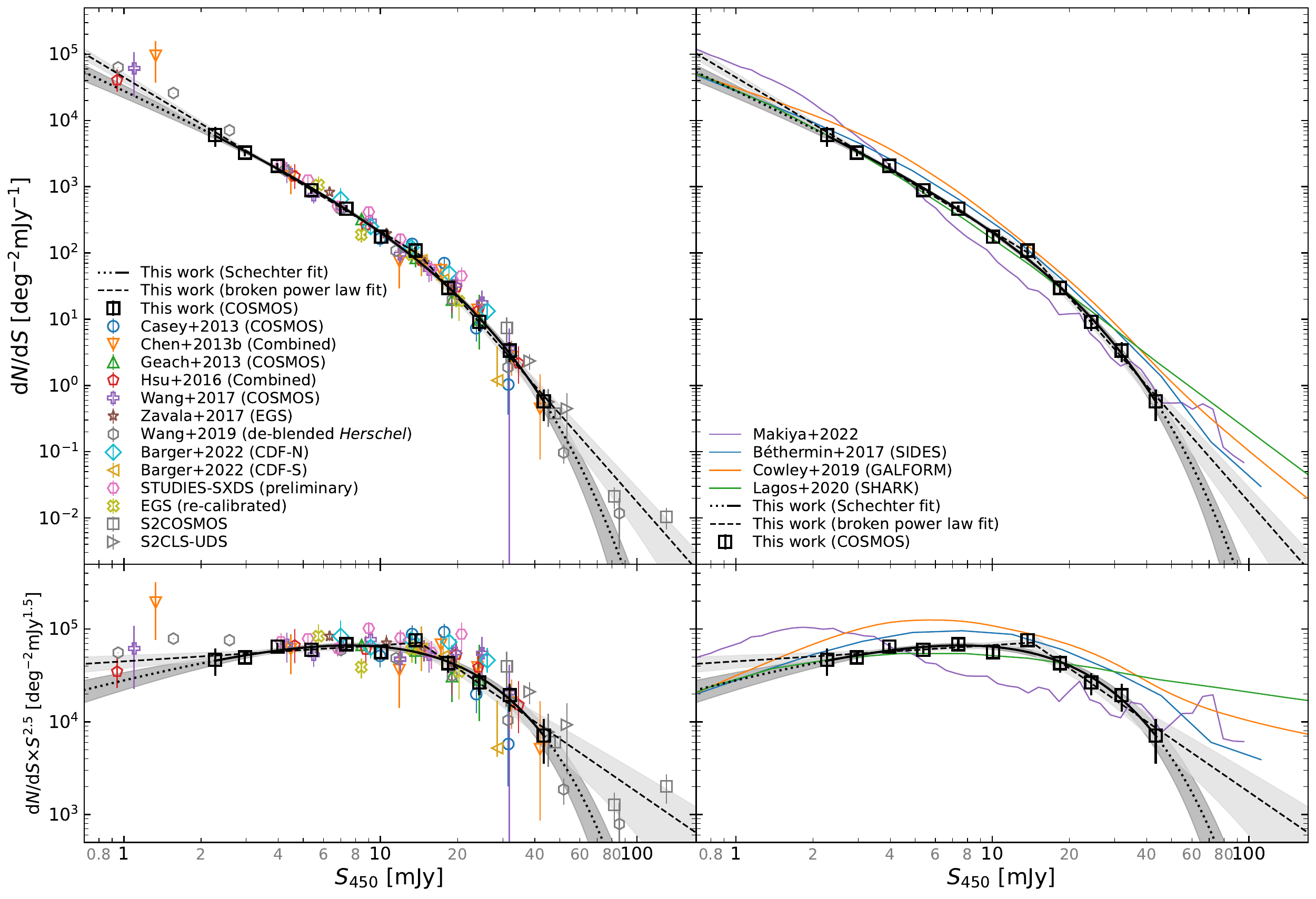}
\caption{Differential number counts (top) and Euclidean-normalized number counts (bottom) at 450~$\micron$. The black solid curve shows the best-fit Schechter function. The extrapolation is shown as a dotted curve, with the light gray band representing the uncertainty. The black dashed curve shows the best-fit broken power law, and the dark gray band indicates the uncertainty. For comparison, we plot the observational results (left) and the model predictions (right) from the literature. In the left panels, we show the SCUBA-2 450-$\micron$ observations and the deblended \emph{Herschel} counts converted from 500~$\micron$ to 450~$\micron$ with a scaling factor \citep{wangMultiwavelengthDeblendedHerschel2019}. Note that some of the counts may include observations from earlier work and are therefore not completely independent.}
\label{fig:450_counts}
\end{figure*}

At the $>30$~mJy bright end, the counts are mostly contributed by the observations of \citet{caseyCharacterizationSCUBA24502013} and the sample size is small (5 sources). To increase the sample size, we reduced the 450-$\micron$ ``PONG'' data from the S2CLS-UDS \citep{geachSCUBA2CosmologyLegacy2017} and the SCUBA-2 COSMOS survey \citep[S2COSMOS,][]{simpsonEastAsianObservatory2019} and derived the bright-end counts from these two wide-field surveys. Because these observations are wide-field and conducted in band-2 weather conditions, which have lower atmospheric transmission, the resulting maps are quite shallow. The areas and central rms noise levels of these fields are 0.67~deg$^2$ and 11~mJy~beam$^{-1}$ for S2CLS-UDS, and 3.14~deg$^2$ and 8~mJy~beam$^{-1}$ for S2COSMOS.
The consequence of such wide and shallow maps is that the detections will be sparse compared to the spurious sources. 
We therefore set a higher S/N cut of 4 for source extraction and number count derivation. The results are presented in Figure~\ref{fig:450_counts} as gray right-pointing triangles (S2CLS-UDS) and gray squares (S2COSMOS).

\subsubsection{Comparisons with Observations}

\begin{figure*}
\epsscale{1.17}
\plotone{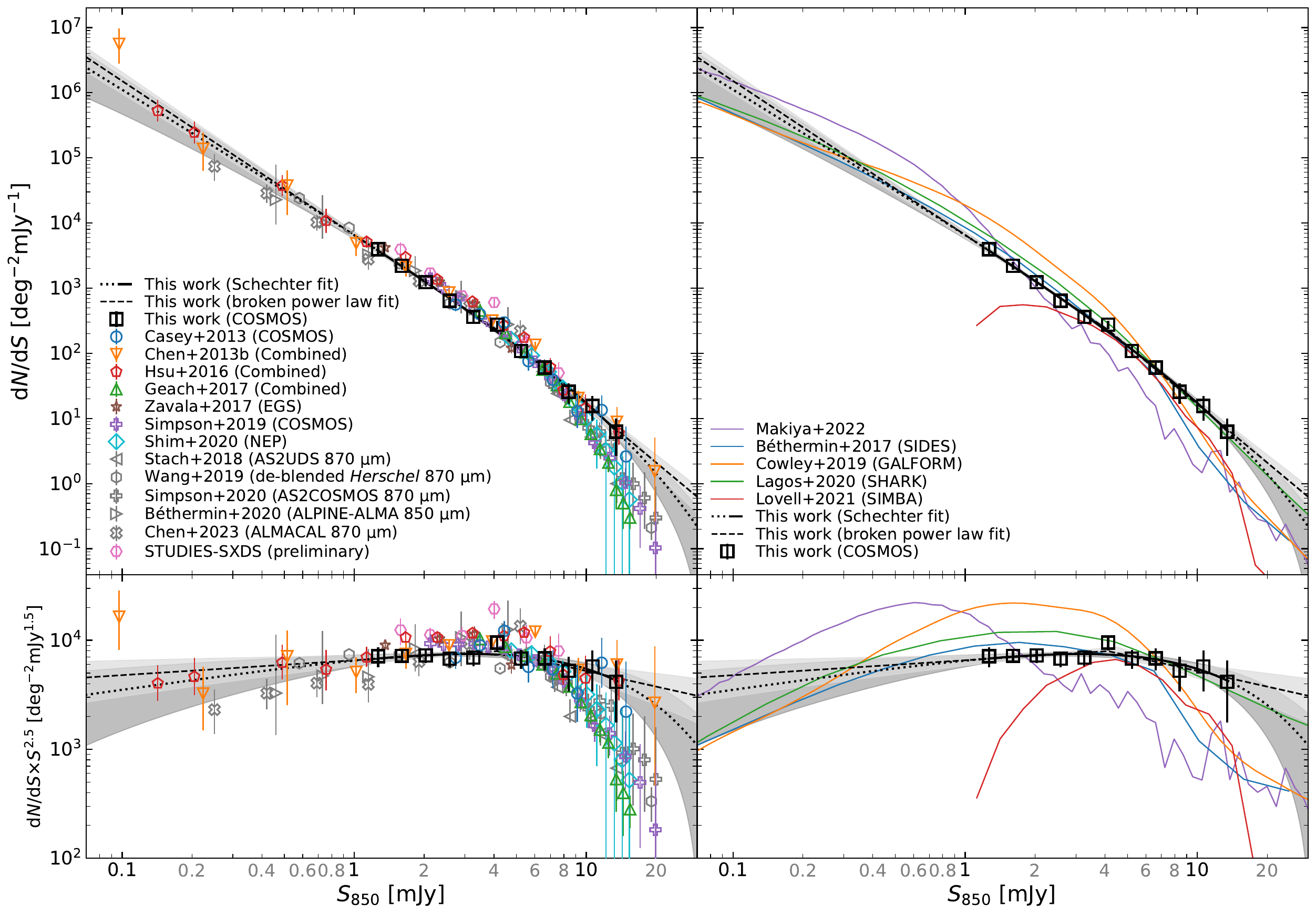}
\caption{Differential number counts (top) and Euclidean-normalized number counts (bottom) at 850~$\micron$. The black solid curve shows the best-fit Schechter function. The extrapolation is shown as a dotted curve, with the light gray band representing the uncertainty. The black dashed curve shows the best-fit broken power law, and the dark gray band indicates the uncertainty. Observational results (left) and model predictions (right) from the literature are plotted for comparison.}
\label{fig:850_counts}
\end{figure*}

We present the comparison of the 450-$\micron$ number counts from this work and the literature counts in Figure~\ref{fig:450_counts}, and the 850-$\micron$ counts in Figure~\ref{fig:850_counts}. The following comparisons of the number counts are made directly based on the published counts, and are independent of the adopted function form for the fitting in each work.
The 850-$\micron$ counts include the results from \citet{caseyCharacterizationSCUBA24502013}, \citet{chenResolvingCosmicFarinfrared2013}, \citet{hsuHawaiiSCUBA2Lensing2016}, \citet{geachSCUBA2CosmologyLegacy2017}, \citet{zavalaSCUBA2CosmologyLegacy2017}, \citet{simpsonEastAsianObservatory2019}, and additional model predictions from \citet{lovellReproducingSubmillimetreGalaxy2021}.
Our bright-end number counts ($>6$ mJy) at 850~$\micron$ exceed those previously reported in the literature.
To validate the excess bright-end counts at 850~$\micron$, we cross-matched our 14 $S_{\mathrm{850}}>8 \;\mathrm{mJy}$ sources with the S2COSMOS catalog \citep{simpsonEastAsianObservatory2019}, which encompasses the entire STUDIES region within its larger survey footprint, and also with the ALMA 870-$\micron$ follow-up observations of S2COSMOS \citep[AS2COSMOS,][]{simpsonALMASurveyBrightest2020}. All 14 sources are detected by S2COSMOS with a flux ratio $S_{\mathrm{STUDIES}}/S_{\mathrm{S2COSMOS}}$ of 1.05$\pm$0.05. There are 13 sources in the AS2COSMOS footprint, and all are recovered with a flux ratio $S_{\mathrm{STUDIES}}/S_{\mathrm{AS2COSMOS}}$ of 1.12$\pm$0.08 compared to the brightest counterparts. Therefore, the existence of a larger number of bright 850~$\micron$ sources as seen in the counts cannot be explained by spurious sources nor by flux boosting. On the other hand, the excess could be attributed to an over-density at $z=2.47$ reported by \citet{caseyMassiveDistantProtocluster2015}.
In addition, the excess also makes the $S_{0}$ (14.5$\pm$9.0 mJy) in the best-fit Schechter function different, though not significantly given the large uncertainty, from other works that are typically in the range of 2--5 mJy \citep[e.g.,][]{hsuHawaiiSCUBA2Lensing2016,geachSCUBA2CosmologyLegacy2017,simpsonEastAsianObservatory2019}.

One of the goal in this work is to go beyond the 850-$\micron$ confusion limit and detect additional sources through 450-$\micron$ observations. By comparing the cumulative number counts at 450 and 850~$\micron$ listed in Tables~\ref{tab:450_counts} and \ref{tab:850_counts} above the confusion limits (see Appendix~\ref{apd:confusion_limit} for more details), we confirmed that confusion-limited 450-$\micron$ observations enables the detection of fainter sources. These fainter sources are expected to lie at lower redshifts, based on the fact that the negative \emph{K}-correction at 450~$\micron$ is only effective out to $z \sim$ 3--4 \citep[e.g.,][]{caseyAreDustyGalaxies2014}.

The SCUBA-2 counts presented in this paper have been corrected for inconsistent FCFs (see Appendix~\ref{apd:fcf_corr_factor}). Besides the counts from the SCUBA-2 observations, we also include the deblended \emph{Herschel} counts at both wavelengths from \citet{wangMultiwavelengthDeblendedHerschel2019}. \citet{wangMultiwavelengthDeblendedHerschel2019} applied a scaling factor of $S_{\mathrm{450}}/S_{\mathrm{500}}=0.86$ to convert the de-blended 500-$\micron$ counts to 450~$\micron$, and they used the 870-$\micron$ flux densities predicted by the SED fitting tool Code Investigating GALaxy Emission \citep[CIGALE,][]{burgarellaStarFormationDust2005,nollAnalysisGalaxySpectral2009,serraCIGALEMCGalaxyParameter2011,boquienCIGALEPythonCode2019} to construct the 870-$\micron$ number counts. We also include the 870-$\micron$ counts from interferometric observations \citep{stachALMASurveySCUBA22018,betherminALPINEALMACIISurvey2020,simpsonALMASurveyBrightest2020,chenALMACALIXMultiband2023}, which suffer less from source blending. 

Compared to other blank-field SCUBA-2 observations, our confusion-limited 450-$\micron$ number counts are about twice as deep, reaching approximately 2~mJy. At this flux limit we find no evidence for a faint-end turnover. Moreover, both the extrapolations of the best-fit Schechter and broken power-law functions from our counts agree with the counts derived from the lensing-cluster fields \citep{hsuHawaiiSCUBA2Lensing2016} and fluctuation analysis in the COSMOS field \citep{wangSCUBA2UltraDeep2017} at around 1~mJy, within the error bars.

Around the knee of the counts (approximately 10 to 30~mJy), our counts are consistent with those in the literature, within the error bars. However, the counts derived from our preliminary STUDIES-SXDS map appear to be more overdense than those derived from STUDIES-COSMOS, especially at the bright end of $S_{\mathrm{450}}\sim20$~mJy. This may be due to the small survey area at this flux density range, where the field-to-field variance may be greater. However, compared to the counts averaged from the different fields (Figure~\ref{fig:450_counts_comparison}), the STUDIES-SXDS counts fall within the uncertainties.

At the very bright end of $>30$~mJy, except for the two wide-field surveys, most of the SCUBA-2 450-$\micron$ data points come from the same sources in the COSMOS field. An exception is the brightest count of \citet{chenResolvingCosmicFarinfrared2013}, derived from the two lensing-cluster fields A1689 and A370. However, possibly due to the use of wider bins, the bin centers of the corresponding data points in the follow-up study \citep{hsuHawaiiSCUBA2Lensing2016} are below 30~mJy. Because the counts at $\gtrsim 30$~mJy are contributed by the same sources, we further verified that the results derived by different methods are consistent with each other, except for the counts of \citet{caseyCharacterizationSCUBA24502013}. Those authors derived the correction parameters (e.g., the flux-boosting factor) of each source based on its S/N rather than on its flux density and local noise. This may have resulted in a greater uncertainty in the counts they derived. As a reference, the 450-$\micron$ de-blended \emph{Herschel} counts and the shallow counts of S2COSMOS and S2CLS-UDS generally agree with our bright-end counts and fall between the extrapolations of the Schechter function and the broken power law.

\subsubsection{Field-to-field Variance and Implications from Models}

\begin{figure*}
\epsscale{1.17}
\plotone{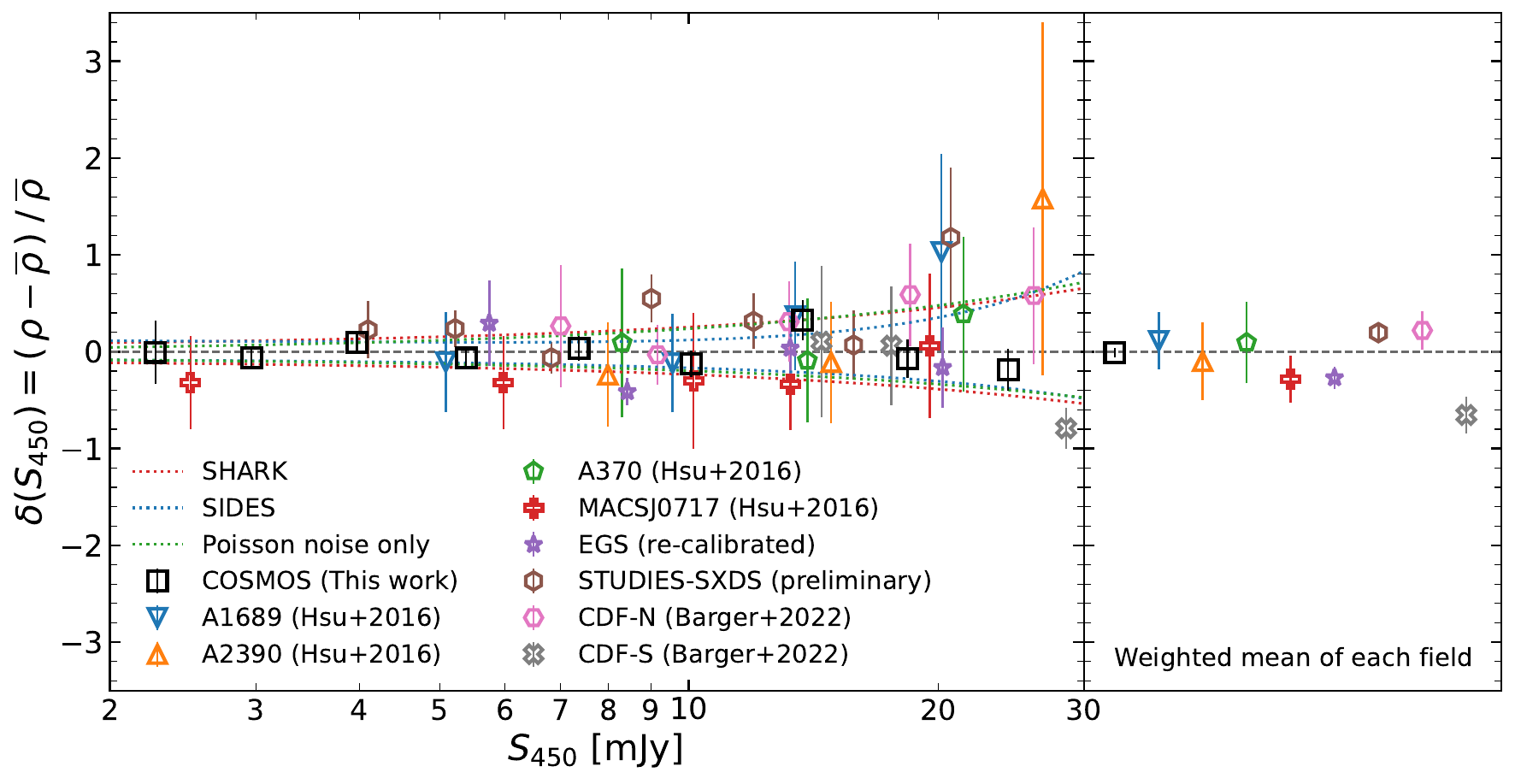}
\caption{Field-to-field variance of 450-$\micron$ counts. The left panel shows the deviations from the mean density at 450~$\micron$ from 2 to 30~mJy. Each field has an area of 110~arcmin$^2$ (approximately 6$\arcmin$ in radius), except for the COSMOS field of this work (12$\arcmin$ in radius). The right panel shows the weighted mean of the $\delta(S_{\mathrm{450}})$ values for each field. The weighted means are generally within 30 percent of the overall mean density, except for the CDF-S field. The dotted curves show the $\sigma_{\mathrm{model}}$ for the SIDES, SHARK, and Poisson-noise-only models. The error bars in the left panel represent the propagated uncertainties on the $\delta(S_{\mathrm{450}})$ values. For the right panel, the error bars denote the errors on the weighted mean.}
\label{fig:450_counts_comparison}
\end{figure*}

Besides comparing specific segments of the counts, we further evaluate the comprehensive variance across fields by employing models as fiducials.
For the field-to-field variance, meaningful field-to-field comparisons can only be made in the flux density range of approximately 2 to 30~mJy, where the counts are best constrained. We adopted these nine independent fields: the STUDIES-COSMOS field of this work; the four lensing-cluster fields (A1689, A2390, A370, and MACSJ0717) from \citet{hsuHawaiiSCUBA2Lensing2016}; the CDF-N and CDF-S fields from \citet{bargerSubmillimeterPerspectiveGOODS2022a}; the EGS field (re-reduced by our pipeline); and the STUDIES-SXDS field. We use the best-fit Schechter function ($N_{0}=$~4626~deg$^{-2}$, $S_{0}=$~11.08~mJy, $\alpha=-$1.81) for the counts of the nine independent fields as the mean density, and then calculate the deviation from the mean density, which can be expressed as $\delta(S_{\mathrm{450}})=(\,\rho-\overline{\rho}\,)\;/\;\overline{\rho}$. In Figure \ref{fig:450_counts_comparison}, we show the quantity $\delta(S_{\mathrm{450}})$ for the nine fields within scales of $R=12\arcmin$ for this work (COSMOS) and $R\sim6\arcmin$ for the others.
The weighted means of $\delta$ over all flux bins within each field (right panel of Figure~\ref{fig:450_counts_comparison}) are generally within 30 percent across the flux range of interest, except for the CDF-S field.

We can further investigate whether the variance seen in Figure~\ref{fig:450_counts_comparison} can be explained by clustering, as seen in the various models, and/or by Poisson errors (i.e., limited sample size). To do this, we performed a $z$-score normalization on the observed data points. The $z$-score at a flux density bin is defined as $z=\delta_{\mathrm{obs}}/\sigma_{\mathrm{model}}$, where $\delta_{\mathrm{obs}}$ is the observed deviation from the mean density, and $\sigma_{\mathrm{model}}$ is the 16th (for negative $\delta_{\mathrm{obs}}$) or 84th (for positive $\delta_{\mathrm{obs}}$) percentile of a sample of $\delta$ measured from a model. We derived the $\sigma_{\mathrm{model}}$ as a function of flux density at the scale of $R\sim6\arcmin$ from \citet[][SIDES]{betherminImpactClusteringAngular2017} and \citet[][SHARK]{lagosPhysicalPropertiesEvolution2020}. To determine if the observed counts are still dominated by Poisson noise, we calculated the $\sigma$ for a scenario in which there is only Poisson noise present. To do this, we created 10,000 simulated images with a radius of 6~arcmin, utilizing the best-fit Schechter function of all observed counts, and adding only Poisson noise and excluding any clustering effects (see the dotted curves in Figure~\ref{fig:450_counts_comparison} for these models). We used seven logarithmic bins between 2 and 30~mJy to make the bin widths similar to the narrowest one of the observed counts. This $\sigma_{\mathrm{model}}$ can be considered as an upper limit, since if the field-to-field variance of the models is identical to that in the real world, the observed variance should not be larger than that of the models because the observed counts are in wider bins. To make the comparisons over a more consistent scale, we divided our $R=12\arcmin$ COSMOS map into four $R=6\arcmin$ pieces.
To assess the field-to-field variance relative to the models, we divided $\delta(S_{\mathrm{450}})$ by $\sigma_{\mathrm{model}}$ to obtain the $z$ scores. The weighted means of the $z$ scores are 1.3$\pm$0.2 (SIDES), 1.0$\pm$0.2 (SHARK), and 1.1$\pm$0.2 (Poisson noise only and no clustering). These results suggest that the observed field-to-field variance is mainly driven by Poisson noise and that there is no strong evidence of clustering at the scale of $R=6\arcmin$ with the current sample size. However, this does not explain the larger observed variance compared with SIDES.

The smaller field-to-field variance in SIDES mentioned above requires a smaller Poisson noise, and this is consistent with the overabundance of sources in SIDES between 2 and 30~mJy, compared to observations (right panels of Figure \ref{fig:450_counts}). This would also require smaller halo masses in SIDES, so that the galaxies in SIDES are more abundant and less strongly clustered. In \citet{limSCUBA2UltraDeep2020a}, the measured halo masses of 450-$\micron$-selected SMGs with $S_{\rm{450}}>4$ mJy at $z=0.5\textrm{--}3$ are $\simeq (2.0\pm0.5)\times 10^{13} h^{-1} \rm M_{\odot}$. From the SIDES catalog using the same flux and redshift selection criteria, assuming a \emph{Planck} cosmology, we found a mean halo mass of $(3.28\pm0.05)\times 10^{12} \,\rm M_{\odot}$. The corresponding halo mass from SIDES is almost 10 times smaller than that in the observations of \citet{limSCUBA2UltraDeep2020a}. This is consistent with the expectation from the overabundance of sources and the larger $z$ score when SIDES is used as the reference model.
It is worth noting that source clustering at the scale of the 450 $\micron$ beam size is not likely to bias the observed counts, based on the analysis by \citet{wangSCUBA2UltraDeep2017}. The authors assessed this effect by placing SIDES sources onto two sets of true-noise maps: one set with the original catalog positions in SIDES (clustered), while the other with random positions (unclustered). They did not find a systematic difference between the counts derived from the two sets.

\begin{figure*}
\epsscale{0.8}
\plotone{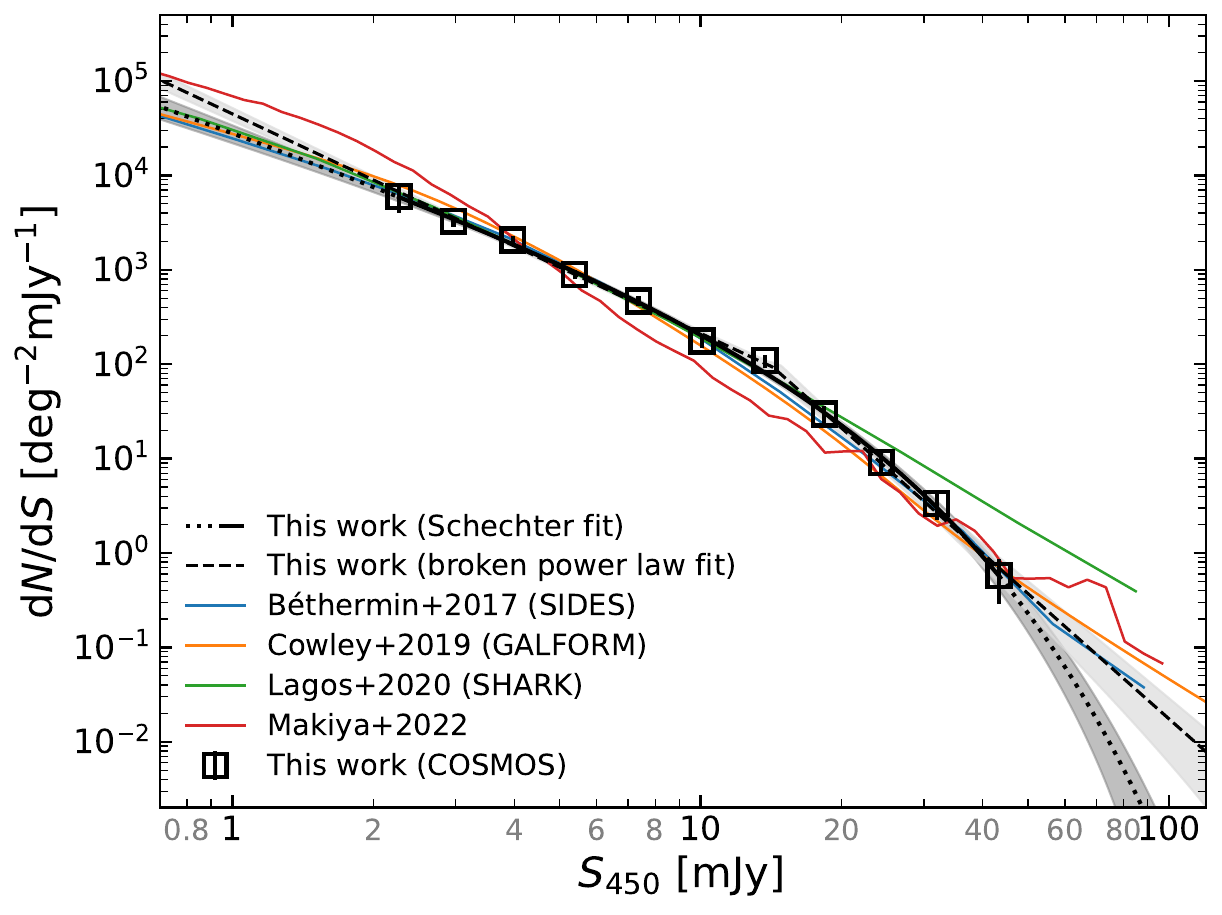}
\caption{Differential number counts at 450~$\micron$ with flux-density adjustments to match the model-predicted counts. Flux-density adjustments of $-$20\% (SIDES), $-$30\% (GALFORM), $-$8\% (SHARK), and 3\% \citep{makiyaCosmicEvolutionGrain2022} are applied to minimize the differences between the predicted and observed counts. Note that if a correction factor of $f$ is applied to $S_{\mathrm{450}}$, a factor of $1/f$ should also be applied to $dN/dS$.}
\label{fig:450_corr_model_counts}
\end{figure*}

Following the above discussion, if the models wrongly assigned an intrinsically less clustered (i.e., less massive) population to be compared against the observed galaxies in our 450-$\micron$ survey, then the models would overpredict the counts and underpredict the clustering. We further examined this possibility.
In the right panel of Figure \ref{fig:450_counts}, we compare our 450-$\micron$ counts with the counts predicted by \citet[][SIDES]{betherminImpactClusteringAngular2017}, \citet[][GALFORM]{cowleyEvolutionUVtommExtragalactic2019}, \citet[][SHARK]{lagosPhysicalPropertiesEvolution2020}, and \citet[][]{makiyaCosmicEvolutionGrain2022}. 
Except for the model of \citet{makiyaCosmicEvolutionGrain2022}, all other models seem to overpredict counts either over the entire flux range of interest, or at least over a significant portion of the flux range.
The predicted counts from SIDES, GALFORM, SHARK, and \citet{makiyaCosmicEvolutionGrain2022} require flux density adjustments of $-$20\%, $-$30\%, $-$8\%, and 3\%, respectively, to minimize the differences between these counts and the observed counts. The results of such adjustments of the model counts are shown in Figure \ref{fig:450_corr_model_counts}. We found that the offsets in the flux densities are not likely to be a consequence of flux calibration problems in the SCUBA-2 observations. Using the FCF measurements provided by \citet{mairsDecadeSCUBA2Comprehensive2021}, we calculated the standard error of the mean of the peak FCF at 450~$\micron$ to be 0.9\% (before June 30, 2018) and 2.7\% (after June 30, 2018). In addition, the systematic uncertainty from the Uranus flux model is 5\% \citep{mairsDecadeSCUBA2Comprehensive2021}, which should be added quadratically to the uncertainty of the FCFs, resulting in total uncertainties of 5.1\% and 5.7\%. These uncertainties are smaller than the observed flux density offsets between most observations and models. Therefore, an overprediction of 450-$\micron$ flux densities of intrinsically less massive galaxies in the models remains a plausible explanation, and this should be tested further.

\subsection{Contribution to the 450-\texorpdfstring{$\mu$}{u}m EBL}

Because we have reached the confusion limit of JCMT SCUBA-2 at 450~$\micron$, we can estimate how much of the EBL at 450~$\micron$ has been (or can be) directly resolved. For the unresolved population, we can infer its nature by extrapolating from our observations.

The \emph{COBE} FIRAS 450-$\micron$ EBL estimated by three groups using different foreground-subtraction methods are 109~Jy~deg$^{-2}$ \citep{pugetTentativeDetectionCosmic1996}, 142~Jy~deg$^{-2}$ \citep{fixsenSpectrumExtragalacticFar1998}, and 150~Jy~deg$^{-2}$ \citep{gispertImplicationsCosmicInfrared2000a}, with large uncertainties of around 30\% or greater. 
We also considered the new determination from \citet{odegardDeterminationCosmicInfrared2019}. They used \emph{COBE} FIRAS data to recalibrate the zero levels and gains of the \emph{Planck} HFI maps released in 2015 and correspondingly obtained better estimates for the 450-$\micron$ EBL. The uncertainty becomes about five times smaller than the previously mentioned $30$\%, and the estimated 450-$\micron$ EBL becomes 134~$\pm$~8~Jy~deg$^{-2}$. This value is very close to the average of the three \emph{COBE} FIRAS measurements. Therefore, we use the value 134~$\pm$~8~Jy~deg$^{-2}$ in our analysis.

\begin{figure*}
\epsscale{1.0}
\plotone{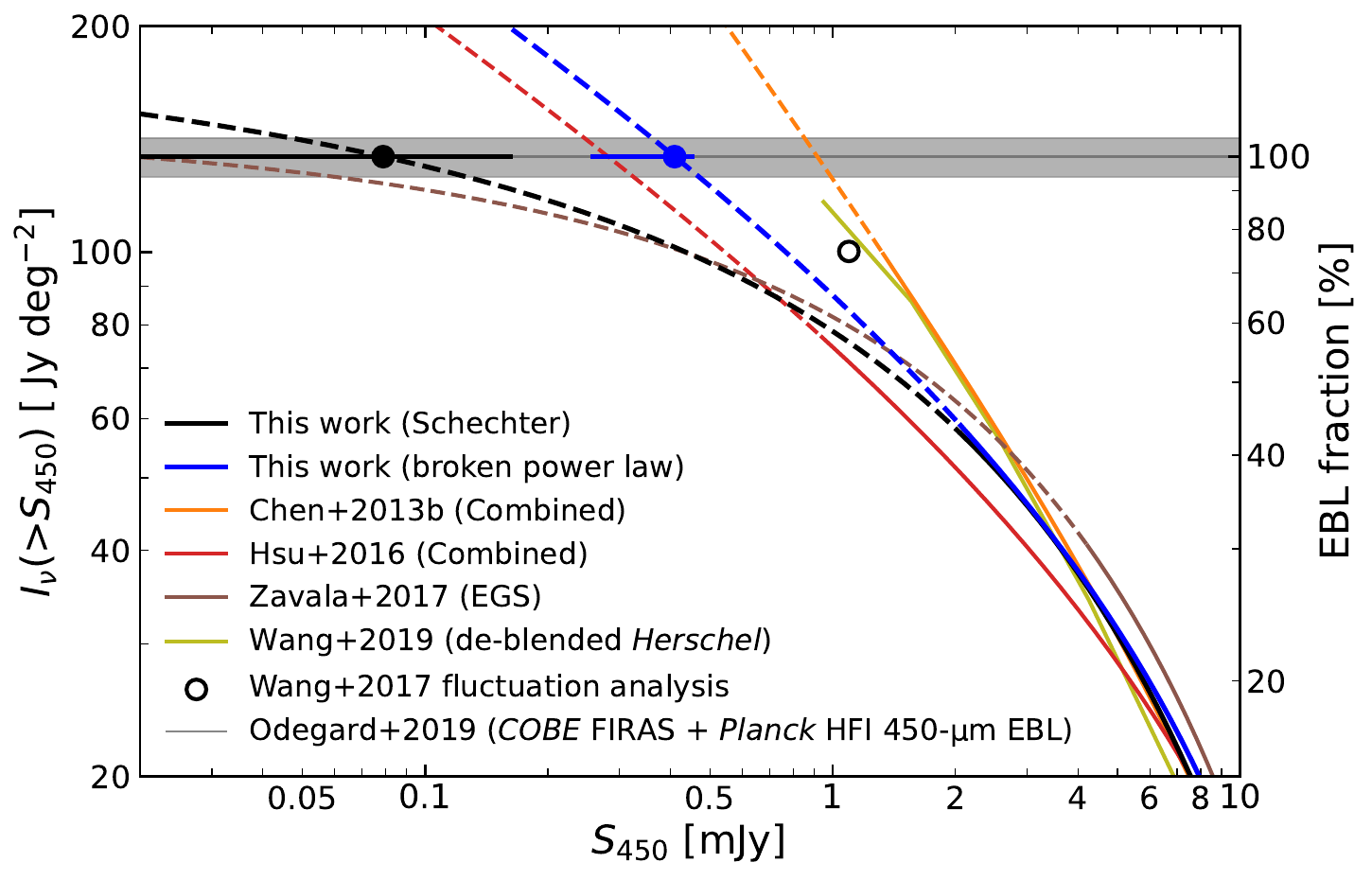}
\caption{Integrated surface brightness at 450~$\micron$. The gray horizontal line and the gray area represent the 450-$\micron$ EBL estimate ($134\pm8$~Jy~deg$^{-2}$) from \citet{odegardDeterminationCosmicInfrared2019}, which combines the results from \emph{COBE} FIRAS and \emph{Planck} HFI; this value is essentially the same as the mean of the previous 450-$\micron$ \emph{COBE} EBL estimates from \citet{pugetTentativeDetectionCosmic1996}, \citet{fixsenSpectrumExtragalacticFar1998}, and \citet{gispertImplicationsCosmicInfrared2000a}. On the right side of the $y$-axis, we use this value for the 100\% resolution level. The solid curves show the results from various SCUBA-2 observations, with the extrapolations shown as the dashed extensions of the curves. We also include the deblended \emph{Herschel} 450-$\micron$ results, which were converted from 500~$\micron$ with a scaling factor of $S_{\rm 450}/S_{\rm500}=0.86$, as the olive curve.
The black empty circle shows the fluctuation analysis result from \citet{wangSCUBA2UltraDeep2017}. The solid circles and horizontal error bars show the flux densities and uncertainties when the extrapolations reach the lower bound (109~Jy~deg$^{-2}$) and the assumed 100\% level (134~Jy~deg$^{-2}$) of the 450-$\micron$ \emph{COBE} EBL estimates. These are the implied detection limits required to fully resolve the EBL.}
\label{fig:450_ebl}
\end{figure*}

\begin{figure}
\epsscale{1.17}
\plotone{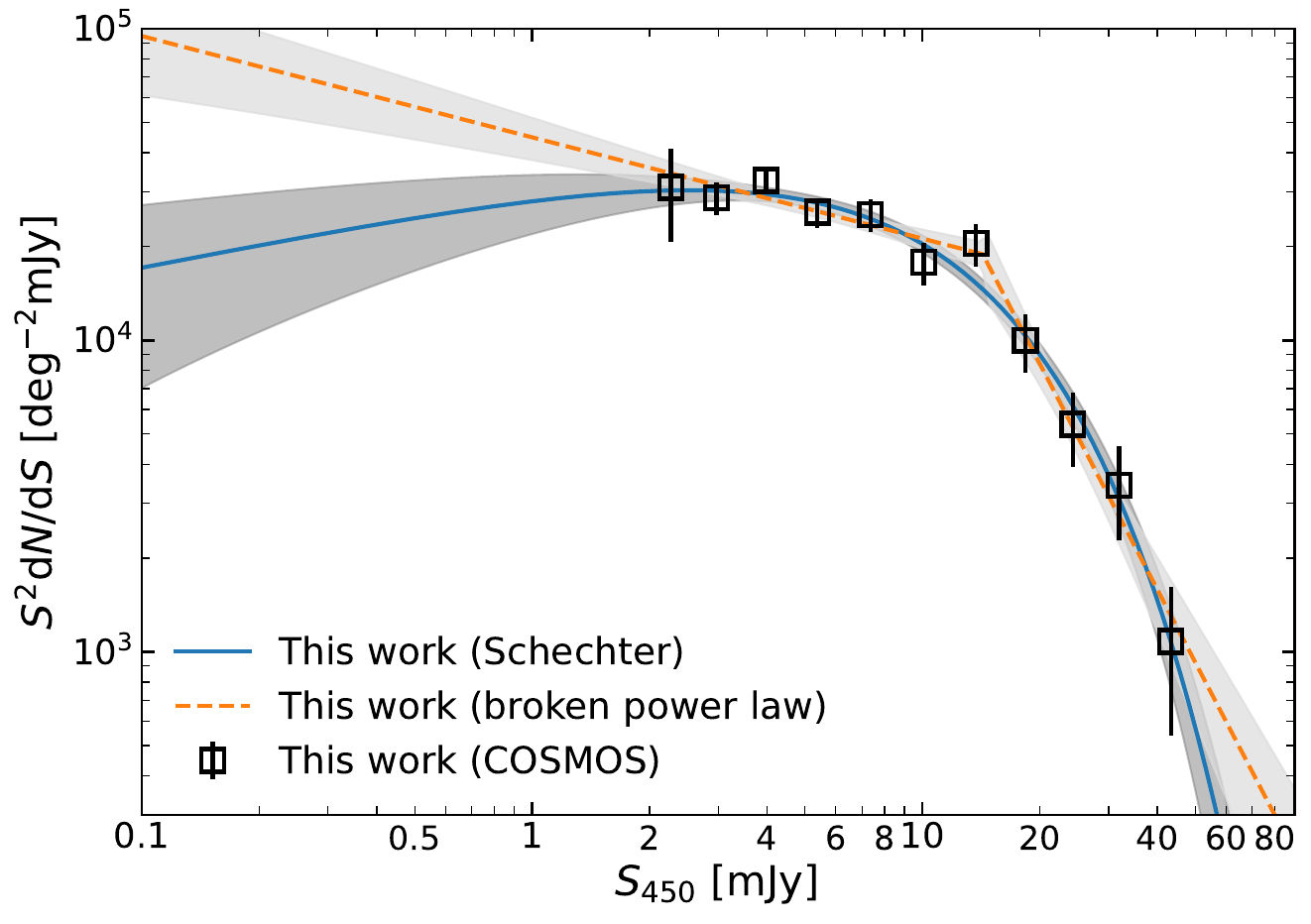}
\caption{Source contributions to the 450-$\micron$ EBL as a function of flux density. The contributions are estimated by applying a weighting of $S^2$ to the differential number counts. The light gray and dark gray shaded regions represent the uncertainties on the contributions from the best-fit broken power law and Schechter functions, respectively. The Schechter fit peaks at $2.5^{+0.9}_{-1.3}$ mJy, indicating the source flux density with the maximum contribution to the EBL.}
\label{fig:450_ebl_contri}
\end{figure}

We integrated the 450-$\micron$ differential number counts using both the best-fit Schechter function and the broken power law. The results are shown in Figure \ref{fig:450_ebl} as black and blue curves, respectively. To estimate the uncertainties, we randomly generated 10,000 $\textit{dN/dS}$ curves from the best-fit parameters along with the covariance matrix. After integrating the generated curves, we can find the values of the 16th and 84th percentiles (i.e., $\pm1\sigma$ levels) as the uncertainties. In Figure \ref{fig:450_ebl}, we show that down to 2.1~mJy (the de-boosted flux density of the faintest $>3.5\sigma$ source), our confusion-limited 450-$\micron$ observations can account for a surface brightness of $57.3^{+1.0}_{-6.2}$~Jy~deg$^{-2}$ (Schechter) and $58.7^{+0.9}_{-8.0}$~Jy~deg$^{-2}$ (broken power law).
These correspond to EBL percentages of (41$\pm$4)\% (Schechter) and $42^{+4}_{-5}\%$ (broken power law). By integrating the extrapolation, it is estimated that the EBL can be 95\% resolved at $0.12^{+0.09}_{-0.12}$~mJy (Schechter) and $0.46^{+0.05}_{-0.17}$~mJy (broken power law) and fully resolved at $0.08^{+0.09}_{-0.08}$~mJy (Schechter) and $0.41^{+0.05}_{-0.16}$~mJy (broken power law). We note that when discussing the full resolution of the EBL, the broken (double) power law requires an additional power law at the faint end to comply with the EBL limit. However, we have no constraints in this faint regime. Therefore, we only take the estimates from the broken power law as upper limits.
From Figure \ref{fig:450_ebl} we can see that the  lensing-cluster counts from \citet{hsuHawaiiSCUBA2Lensing2016} reached a deeper limit of approximately 0.9~mJy. Their integrated surface brightness at this flux level is similar to our Schechter function extrapolation. Their extrapolation falls between our Schechter function extrapolation and broken power-law extrapolation at fainter flux levels. Recently, \citet{hsuFullResolution4502024} reported the full resolution of the 450-$\micron$ EBL at $\sim0.1$~mJy using the latest SCUBA-2 observations of lensing clusters. This is in excellent agreement with our Schechter extrapolation.
These results show that the detection limit required for a full resolution of the EBL depends on the count slope at $<1$~mJy, and that neither our confusion-limited blank-field counts nor the counts derived from lensing-cluster fields can constrain this. In addition, we can assess the source contributions to the EBL as a function of flux density by adopting the corresponding weighting of $S^2 dN/dS$ used in \citet{vernstromDeepDiffuseExtragalactic2015}. Figure~\ref{fig:450_ebl_contri} shows that the Schechter fit presents a maximum contribution at $2.5^{+0.9}_{-1.3}$ mJy. Again, the broken power-law fit needs an additional break so that it does not exceed the EBL.

Finally, we can ask what kind of galaxies correspond to the above flux densities required for full resolution of the 450-$\micron$ EBL. The higher flux density of $S_{\mathrm{450}}=0.41$~mJy (broken power-law extrapolation) corresponds to infrared luminosities of $\rm L_{\rm IR} = 5.7\times10^{10}~L_{\odot}$ at $z=1$ and $\rm L_{\rm IR} = 9.8\times10^{10}~L_{\odot}$ at $z=2$, assuming the average ALESS SMG SED \citep{cunhaALMASurveySubmillimeter2015}. These correspond to SFRs of 8~$\rm M_{\odot}\;\mathrm{yr}^{-1}$ at $z=1$ and 15~$\rm M_{\odot}\;\mathrm{yr}^{-1}$ at $z=2$. On the other hand, for the lower flux density of $S_{\mathrm{450}}=0.08$~mJy (Schechter function extrapolation), the corresponding infrared luminosity and SFR at $z=1$ are $5.7\times10^{9}~\rm L_{\odot}$ and 0.9 $\rm M_{\odot}\;\mathrm{yr}^{-1}$, and $1.6\times10^{10}~\rm L_{\odot}$ and 2.4 $\rm M_{\odot}\;\mathrm{yr}^{-1}$ at $z=2$, respectively, assuming a normal galaxy SED template from \citet{schreiberDustTemperatureMidtototal2018} that does not incorporate any $T_{\rm d}\textrm{--} L_{\rm IR}$ relation, but instead has $T_{\rm d}$ evolving with redshift. Both of the above SFRs fall into the range of star-forming galaxies detected in deep optical surveys. 
Furthermore, we can compare the number densities of such faint 450-$\micron$ sources with those detected in deep optical surveys. The cumulative 450-$\micron$ counts at the flux density limits required to fully resolve the EBL are $2.5 \times 10^5$~deg$^{-2}$ (Schechter) and $1.1 \times 10^5$~deg$^{-2}$ (broken power law). At $z=1\;(2)$, a less obscured star-forming galaxy with $A_V<1$ and an extinction-corrected SFR of 5 (10)~$M_{\odot}\;\mathrm{yr}^{-1}$ would have an optical magnitude of $R\sim24\;(25)$ \citep[e.g.,][]{weaverCOSMOS2020PanchromaticView2022}. The cumulative counts of optical galaxies with $R=24-25$ are estimated to be in the range of $2.0 \times 10^4$ to $5.6 \times 10^4$~deg$^{-2}$ \citep{smailDeepOpticalGalaxy1995,metcalfeGalaxyNumberCounts2001,capakDeepWideFieldOptical2004,kashikawaSubaruDeepField2004}. This is within factors of 13 (Schechter function) and 6 (broken power law) to the density of faint 450-$\micron$ sources. In a very approximate way, this suggests that a 450-$\micron$ survey needs to detect the dust emission from every faint optical galaxy, down to $R\sim24-25$, to fully account for the 450-$\micron$ EBL.

\section{Summary} \label{sec:summary}
We have presented a confusion-limited SCUBA-2 450-$\micron$ blank-field image in the COSMOS-CANDELS region. The observations were mainly contributed by our JCMT large program, STUDIES, completed in mid-2020, along with archival data in the same field. Our maps at 450~$\micron$ and 850~$\micron$ achieved sensitivities of $\sigma_{\mathrm{450}}=0.59$~mJy and $\sigma_{\mathrm{850}}=0.09$~mJy in the deepest area of each map, which are comparable to or lower than the confusion noise levels of 0.65~mJy and 0.36~mJy, respectively. In the $R=12\arcmin$ deep region, we detected 360 (237) $>4\sigma$ and 479 (314) $>3.5\sigma$ sources at 450 (850)~$\micron$, respectively. We constructed catalogs at these two wavelengths using the $>3.5\sigma$ sources. We present these catalogs in Tables \ref{tab:450_catalog} and \ref{tab:850_catalog}, including the de-boosted flux, completeness, and spurious probability for each source estimated with Monte Carlo simulations. We also make our reduced images publicly available.

We constructed differential number counts at 450~$\micron$ using the $>3.5\sigma$ sources, spanning a wide flux density range, from 2.3~mJy to 43.3~mJy. Our counts are consistent with the counts from previous SCUBA-2 blank-field and lensing-cluster surveys in the literature. Our faint-end ($\sim$1~mJy) extrapolation is in good agreement with the counts derived from the lensing-cluster fields and the fluctuation analysis in the blank field. The observed field-to-field variance at 450~$\micron$ at an $R=6\arcmin$ scale is consistent with Poisson noise, so we do not find evidence of strong clustering at this scale.  On the other hand, through the comparison with models, we find hints of evidence that (some) models may over-predict the 450-$\micron$ flux densities.

With our confusion-limited SCUBA-2 450-$\micron$ map, we have directly resolved (41$\pm$4)\% of the 450-$\micron$ \emph{COBE} EBL at 2.1~mJy. The resolved sources produce an integrated surface brightness of $57.3^{+1.0}_{-6.2}$~Jy~deg$^{-2}$. To fully resolve the 450-$\micron$ EBL of 134~Jy~deg$^{-2}$, estimated by \citet{odegardDeterminationCosmicInfrared2019}, the detection limit should be pushed to
$0.41^{+0.05}_{-0.16}$~mJy (based on a broken power-law extrapolation to our counts) or even $0.08^{+0.09}_{-0.08}$~mJy (Schechter function extrapolation), which may be achievable with extremely deep lensing-cluster observations and next-generation submillimeter facilities with large aperture sizes.

\section{Acknowledgments}
We are grateful to the JCMT/EAO staff for providing observational support, managing the data and surveys, and to the anonymous referee for invaluable feedback that greatly improved the manuscript.
Z.K.G., C.F.L., and W.H.W. acknowledge support from the National Science and Technology Council of Taiwan (NSTC 111-2112-M-001-052-MY3).
C.C.C. acknowledges support from the National Science and Technology Council of Taiwan (NSTC 109-2112-M-001-016-MY3 and 111-2112M-001-045-MY3), as well as Academia Sinica through the Career Development Award (AS-CDA-112-M02).
I.R.S. acknowledges support from STFC (ST/T000244/1).
H.S. acknowledges the support from the National Research Foundation of Korea grant No.2021R1A2C4002725 and No.2022R1A4A3031306, funded by the Korea government (MSIT). 
Y.A. acknowledges support by NSFC grants 12173089.
L.C.H. was supported by the National Science Foundation of China (11721303, 11991052, 12011540375, 12233001) and the China Manned Space Project (CMS-CSST-2021-A04, CMS-CSST-2021-A06).

The James Clerk Maxwell Telescope is operated by the East Asian Observatory on behalf of The National Astronomical Observatory of Japan; Academia Sinica Institute of Astronomy and Astrophysics; the Korea Astronomy and Space Science Institute; the National Astronomical Research Institute of Thailand; Center for Astronomical Mega-Science (as well as the National Key R\&D Program of China with No. 2017YFA0402700). Additional funding support is provided by the Science and Technology Facilities Council of the United Kingdom and participating universities and organizations in the United Kingdom and Canada.
Additional funds for the construction of SCUBA-2 were provided by the Canada Foundation for Innovation.
The James Clerk Maxwell Telescope has historically been operated by the Joint Astronomy Centre on behalf of the Science and Technology Facilities Council of the United Kingdom, the National Research Council of Canada and the Netherlands Organisation for Scientific Research.
The authors wish to recognize and acknowledge the very significant cultural role and reverence that the summit of Maunakea has always had within the indigenous Hawaiian community.  We are most fortunate to have the opportunity to conduct observations from this mountain.
The submillimeter observations used in this work include the STUDIES program \citep[][program code M16AL006 and M17BL009]{wangSCUBA2UltraDeep2017}, S2CLS program \citep[][program codes MJLSC01 and MJLSC02]{geachSCUBA2CosmologyLegacy2013, geachSCUBA2CosmologyLegacy2017,zavalaSCUBA2CosmologyLegacy2017}, S2COSMOS program \citep[][program codes M16AL002 and M17BL006]{simpsonEastAsianObservatory2019}, and the PI program of \citet[][program codes M11BH11A, M12AH11A, and M12BH21A]{caseyCharacterizationSCUBA24502013}.

%

\vspace{5mm}
\facilities{JCMT (SCUBA-2)}


\software{Astropy \citep{astropycollaborationAstropyCommunityPython2013,astropycollaborationAstropyProjectBuilding2018,astropycollaborationAstropyProjectSustaining2022},
Dask \citep{daskdevelopmentteamDaskLibraryDynamic2016},
NumPy \citep{harrisArrayProgrammingNumPy2020},
pandas \citep{reback2020pandas},
PICARD \citep{jennessJCMTScienceArchive2008},
SciPy \citep{virtanenSciPyFundamentalAlgorithms2020},
SMURF \citep{chapinSCUBA2IterativeMapmaking2013},
Starlink \citep{currieStarlinkSoftware20132014}}



\appendix

\section{Confusion limit}\label{apd:confusion_limit}
The confusion limit $S_{\rm c}$ \citep{scheuerStatisticalMethodAnalysing1957,condonConfusionFluxDensityError1974,franceschiniSourceCountsConfusion1982} is a flux limit below which the detected peaks become less meaningful due to source blending. One rule of thumb for the source density criterion of the confusion limit is one source per 30 beams. The confusion limit can be estimated using the following criterion:
\begin{equation}
    \Omega_{\mathrm{b}} \int_{S_{c}}^{S_{\mathrm{max}}}{\frac{dN}{dS}dS}=\frac{1}{30},
\end{equation}
where $\Omega_{\mathrm{b}}$ is the SCUBA-2 beam area \citep[104~arcsec$^2$ for 450~$\micron$ and 228~arcsec$^2$ for 850~$\micron$,][]{dempseySCUBA2OnskyCalibration2013}, $S_{\mathrm{max}}$ is the de-boosted flux density of the brightest source (45.2~mJy for 450~$\micron$ and 17.0~mJy for 850~$\micron$), $dN/dS$ is the differential number counts, and 1/30 means one source per 30 beams. The estimated confusion limits are 4.4~mJy at 450~$\micron$ and 2.0~mJy at 850~$\micron$. The confusion limits for different source density criteria are listed in Table \ref{tab:confusion_limit}.
We note that our estimated confusion limits are higher than the $\sim 2$~mJy value at 450~$\micron$ estimated by \citet{chenFaintSubmillimeterGalaxy2013} and the 1.68~mJy value at 850~$\micron$ estimated by \citet{cowieSubmillimeterPerspectiveGOODS2017}. This is because they estimated the confusion limits based on smaller beam areas and differential counts rather than cumulative counts (i.e., the integral of the differential counts). If we estimate the confusion limits using differential counts and the same beam areas, the estimated confusion limits would be 2.2~mJy at 450~$\micron$ and 1.5~mJy at 850~$\micron$.  
Furthermore, we can use the confusion limits to infer that the noise level required to detect 450-$\micron$ sources at $>3.5\sigma$ ($>4\sigma$) is 1.26~mJy (1.10~mJy). We note that the faintest sources (2.1~mJy for 450~$\micron$ and 1.3~mJy for 850~$\micron$) in our catalogs correspond to roughly one source per 10 beams.
Although some of our detected sources are below these limits, we still use them in the analyses while considering completeness and spurious detection probabilities. We recommend that users of our final catalogs select sources based on completeness and spurious probability, rather than simply considering the confusion limits. 

\begin{deluxetable}{ccc}
\tablecaption{\label{tab:confusion_limit}Measurements of confusion limit}
\tablehead{\colhead{} & \colhead{450 $\micron$} & \colhead{850 $\micron$}\\
\colhead{(No. of beams)$^{-1}$} & \colhead{(mJy)} & \colhead{(mJy)}} 
\startdata
10 & 2.0 & 1.0 \\
20 & 3.4 & 1.6 \\
30 & 4.4 & 2.0 \\
40 & 5.3 & 2.4 \\
50 & 6.0 & 2.8
\enddata
\tablecomments{The rule-of-thumb source density for the confusion limit is one source per 30 beams.}
\end{deluxetable}

\section{FCF correction factor}\label{apd:fcf_corr_factor}
To compare the various counts in Section \ref{subsec:count_comparison} on a fair basis, we need to consider the flux conversion factors (FCFs) adopted by each team. \citet{mairsDecadeSCUBA2Comprehensive2021} found that the FCF varies when the line-of-sight opacity changes. Moreover, the FCF changed after the update of the SCUBA-2 thermal filter stack in 2016 and after the secondary mirror repair in 2018. The source flux density calibrated by the new standard FCFs derived by \citet{mairsDecadeSCUBA2Comprehensive2021} is a few to $10$ percent higher than the previous standard FCFs given in \citet{dempseySCUBA2OnskyCalibration2013}, which was adopted by most of the previous SCUBA-2 observations. To adjust the previous counts to the new standard FCFs, we examined the FCFs adopted in the literature, the factor to compensate for the flux loss that occurs during the data reduction and calibration processes, and the factor used to correct the extinction relation from an older version to the new one. These values are listed in Table \ref{tab:fcf_factors}. The flux densities of the counts shown in Figures \ref{fig:450_counts} and \ref{fig:850_counts} has been corrected by applying the factor $f_\mathrm{total}$, which can be calculated by
\begin{equation}\label{eq:fcf_corr}
    \left(\frac{\mathrm{FCF}_\mathrm{new}}{\mathrm{FCF}_\mathrm{adopted}}\right) \times \left(\frac{f_\mathrm{loss,this\,work}}{f_\mathrm{loss}}\right) \times f_\mathrm{ext},
\end{equation}
where $\mathrm{FCF}_\mathrm{new}$ is the new FCF published by \citet{mairsDecadeSCUBA2Comprehensive2021}, $\mathrm{FCF}_\mathrm{adopted}$ is the FCF adopted in the literature, $f_\mathrm{loss}$ and $f_\mathrm{loss,this\,work}$ are the flux-loss compensation factors adopted in the literature and in this work (Section \ref{data_reduction}), respectively, and $f_\mathrm{ext}$ is the factor used to correct the extinction relation from an older version to the new one.

\begin{figure*}
\epsscale{1.17}
\plottwo{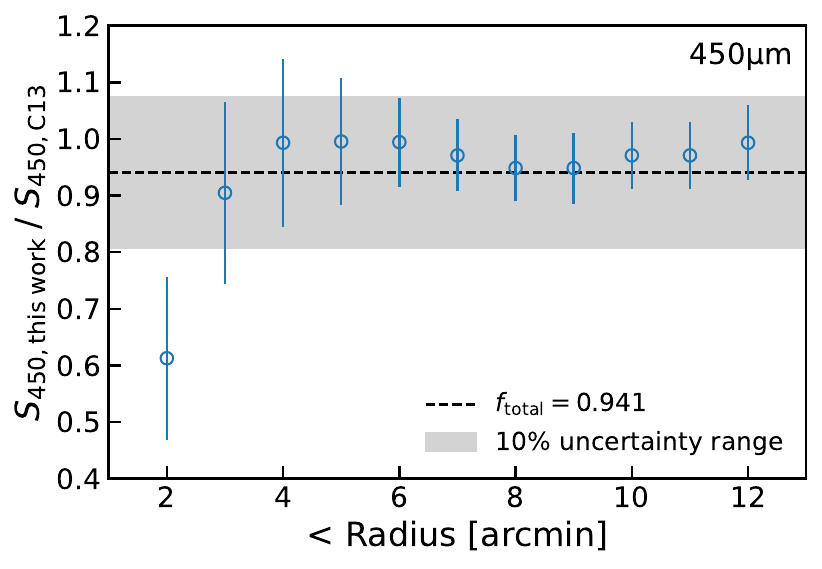}{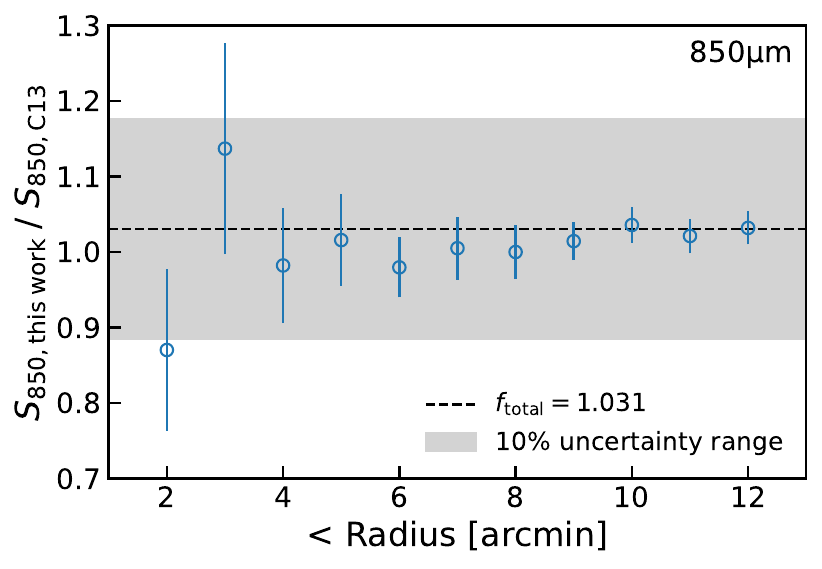}
\caption{Flux ratios of sources in the C13 map and ours at different radius cuts. The dashed horizontal line marks the FCF-correction factor for C13. The gray region is the $\pm$10\% uncertainty range of the data calibration. The results are consistent with the factors within the error bars for both 450 and 850~$\micron$, except at radius $<2$~arcmin.}
\label{fig:c13_factor}
\end{figure*}

One dataset that allows us to validate this flux adjustment is the wide and shallow 450-$\micron$ catalog published by \citet{caseyCharacterizationSCUBA24502013}, which fully overlaps with our map. We used the catalog therein to confirm that the $f_\mathrm{total}$ adjustment factor indeed leads to consistent fluxes. We verified the FCF-correction factors by comparing the peak values of the sources detected in the map of \citet[][hereafter C13]{caseyCharacterizationSCUBA24502013} with ours. First, we found the peaks in our S/N map with a threshold of 4 for a secure selection. Then we used these source positions to find corresponding pixels in the flux map of C13. To ensure that the sample is of good quality, we only selected sources $>4\sigma$ in both maps. In Figure \ref{fig:c13_factor} we show that the sigma-clipped median flux ratios with bootstrapped errors at different radius cuts are consistent with the factors listed in Table \ref{tab:fcf_factors} for both 450 and 850~$\micron$, within the 10\% reduction uncertainty. If we ignore the large 10\% uncertainty of the data calibration, the factors are still consistent with the results within the error bars. This comparison allows us to conclude that the correction with Eq. \ref{eq:fcf_corr} can lead to fair comparisons between the various counts.

\begin{deluxetable*}{cccccc|ccccc}
\tablecaption{\label{tab:fcf_factors} Factors for correcting flux densities in the literature to our standard.}

\tablehead{
 & \multicolumn{5}{c}{450 $\micron$} & \multicolumn{5}{c}{850 $\micron$} \\
\colhead{Literature} & \colhead{FCF$_\mathrm{adopted}$} & \colhead{FCF$_\mathrm{new}$} & \colhead{$f_\mathrm{loss}$} & \colhead{$f_\mathrm{ext}$} & \colhead{$f_\mathrm{total}$} & \colhead{FCF$_\mathrm{adopted}$} & \colhead{FCF$_\mathrm{new}$} & \colhead{$f_\mathrm{loss}$} & \colhead{$f_\mathrm{ext}$} & \colhead{$f_\mathrm{total}$}}

\startdata
  \citet{caseyCharacterizationSCUBA24502013} &   606   & 531     &   1.000  &   1.022 &     0.941 &   556   &   525   &    1.000 &   0.985 &  1.031   \\
  \citet{chenResolvingCosmicFarinfrared2013} &   491   & 531 &   1.100  &   1.022 &     1.056 &   537   &   525   &    1.100 &   0.985 &  0.971   \\
  \citet{geachSCUBA2CosmologyLegacy2013} &   491   & 531     &   1.100  &   1.022 &     1.056 & \nodata & \nodata & \nodata & \nodata &  \nodata \\
  \citet{hsuHawaiiSCUBA2Lensing2016}$^\dag$ &   491   & 531     &   1.160  &   1.022 &     1.001 &   537   &   525   &    1.200 &   0.985 &  0.890   \\
 \citet{geachSCUBA2CosmologyLegacy2017} & \nodata & \nodata & \nodata & \nodata &   \nodata &   537   &   525   &    1.100 &   1.010 &  0.996   \\
 \citet{wangSCUBA2UltraDeep2017} &   490   & 531     &   1.062  &   1.022 &     1.096 & \nodata & \nodata & \nodata & \nodata &  \nodata \\
 \citet{zavalaSCUBA2CosmologyLegacy2017} &   491   & 531     &   1.100  &   1.022 &     1.056 &   537   &   525   &    1.100 &   0.985 &  0.971   \\
 \citet{simpsonEastAsianObservatory2019} & \nodata & \nodata & \nodata & \nodata &   \nodata &   537   &   525   &    1.130 &   1.010 &  0.969   \\
 \citet{shimNEPSC2NorthEcliptic2020}$^\ddag$ & \nodata & \nodata & \nodata & \nodata &   \nodata & \nodata & \nodata &    1.050 & \nodata &  1.056   \\
 \citet{bargerSubmillimeterPerspectiveGOODS2022a} &   491   & 531     &   1.100  &   1.022 &     1.056 & \nodata & \nodata & \nodata & \nodata & \nodata
\enddata

\tablecomments{The FCFs are in Jy~beam$^{-1}$~pW$^{-1}$ units.\\
FCF$_\mathrm{adopted}$ is the FCF adopted in the literature.\\
FCF$_\mathrm{new}$ is the new FCF given by \citet{mairsDecadeSCUBA2Comprehensive2021}.\\
$f_\mathrm{loss}$ is the adopted factor to compensate for the flux loss that occurs during the data reduction and calibration processes.\\
$f_\mathrm{ext}$ is used to correct the extinction relation from an older version to the new one.\\
$f_\mathrm{total}$ can be calculated by $(\mathrm{FCF}_\mathrm{new} / \mathrm{FCF}_\mathrm{adopted}) \times (f_\mathrm{loss,this\,work} / f_\mathrm{loss}) \times f_\mathrm{ext}$, where $f_\mathrm{loss,this\,work}$ is 1.051 for 450~$\micron$ and 1.109 for 850~$\micron$.\\
$\dag$: \citet{hsuHawaiiSCUBA2Lensing2016} derived the FCF from the calibrators with the ``blank-field'' configuration file and found the values to be 16\% and 20\% higher than the standard values at 450~$\micron$ and 850~$\micron$, respectively. They used these values for the calibration without additional flux-loss compensation. Therefore, the $f_\mathrm{loss}$ here has a different meaning than the others.\\
$\ddag$: \citet{shimNEPSC2NorthEcliptic2020} calibrated their data using the pre-release method, which has since been finalized and published by \citet{mairsDecadeSCUBA2Comprehensive2021}. The calibration factors, including FCFs and $f_\mathrm{ext}$, derived from the pre-release method are identical to the ones we used. Therefore, $f_\mathrm{total}$ simplifies to $f_\mathrm{loss,this\,work} / f_\mathrm{loss}$ in this case.}
\end{deluxetable*}

\bibliography{submm}{}
\bibliographystyle{aasjournal}


\end{CJK*}
\end{document}